\RequirePackage{lineno}

\documentclass[aps,prd,twocolumn,superscriptaddress,showpacs,floatfix,nofootinbib]{revtex4}%% For Scientific Linux
\usepackage{comment}
\usepackage{graphicx}
\usepackage[dvips]{color} 
\usepackage{bm}

\newcommand{\MET}{ $/\!\!\!\!E_{T}$} %% CDF recommended ? 

\newcommand{\invfb}{\mbox{fb}^{-1}}

\newcommand{\pythia}{{\sc Pythia} }

\newcommand{\tev}{\mathrm{TeV}}% this is for the equasion
\newcommand{\gevcc}{\mathrm{GeV}/c^2}% this is for the equasion
\begin{document}
\vspace{2cm}
%\preprint{CDF/PUB/EXOTIC/CDFR/10752}
%\preprint{PRD draft version 3.3}

%% 7.5/fb paper
\title{Search for the standard model Higgs boson produced in association with a $W^{\pm}$ boson with 7.5 fb$^{-1}$ integrated luminosity at CDF}
\affiliation{Institute of Physics, Academia Sinica, Taipei, Taiwan 11529, Republic of China}
\affiliation{Argonne National Laboratory, Argonne, Illinois 60439, USA}
\affiliation{University of Athens, 157 71 Athens, Greece}
\affiliation{Institut de Fisica d'Altes Energies, ICREA, Universitat Autonoma de Barcelona, E-08193, Bellaterra (Barcelona), Spain}
\affiliation{Baylor University, Waco, Texas 76798, USA}
\affiliation{Istituto Nazionale di Fisica Nucleare Bologna, $^{ee}$University of Bologna, I-40127 Bologna, Italy}
\affiliation{University of California, Davis, Davis, California 95616, USA}
\affiliation{University of California, Los Angeles, Los Angeles, California 90024, USA}
\affiliation{Instituto de Fisica de Cantabria, CSIC-University of Cantabria, 39005 Santander, Spain}
\affiliation{Carnegie Mellon University, Pittsburgh, Pennsylvania 15213, USA}
\affiliation{Enrico Fermi Institute, University of Chicago, Chicago, Illinois 60637, USA}
\affiliation{Comenius University, 842 48 Bratislava, Slovakia; Institute of Experimental Physics, 040 01 Kosice, Slovakia}
\affiliation{Joint Institute for Nuclear Research, RU-141980 Dubna, Russia}
\affiliation{Duke University, Durham, North Carolina 27708, USA}
\affiliation{Fermi National Accelerator Laboratory, Batavia, Illinois 60510, USA}
\affiliation{University of Florida, Gainesville, Florida 32611, USA}
\affiliation{Laboratori Nazionali di Frascati, Istituto Nazionale di Fisica Nucleare, I-00044 Frascati, Italy}
\affiliation{University of Geneva, CH-1211 Geneva 4, Switzerland}
\affiliation{Glasgow University, Glasgow G12 8QQ, United Kingdom}
\affiliation{Harvard University, Cambridge, Massachusetts 02138, USA}
\affiliation{Division of High Energy Physics, Department of Physics, University of Helsinki and Helsinki Institute of Physics, FIN-00014, Helsinki, Finland}
\affiliation{University of Illinois, Urbana, Illinois 61801, USA}
\affiliation{The Johns Hopkins University, Baltimore, Maryland 21218, USA}
\affiliation{Institut f\"{u}r Experimentelle Kernphysik, Karlsruhe Institute of Technology, D-76131 Karlsruhe, Germany}
\affiliation{Center for High Energy Physics: Kyungpook National University, Daegu 702-701, Korea; Seoul National University, Seoul 151-742, Korea; Sungkyunkwan University, Suwon 440-746, Korea; Korea Institute of Science and Technology Information, Daejeon 305-806, Korea; Chonnam National University, Gwangju 500-757, Korea; Chonbuk National University, Jeonju 561-756, Korea}
\affiliation{Ernest Orlando Lawrence Berkeley National Laboratory, Berkeley, California 94720, USA}
\affiliation{University of Liverpool, Liverpool L69 7ZE, United Kingdom}
\affiliation{University College London, London WC1E 6BT, United Kingdom}
\affiliation{Centro de Investigaciones Energeticas Medioambientales y Tecnologicas, E-28040 Madrid, Spain}
\affiliation{Massachusetts Institute of Technology, Cambridge, Massachusetts 02139, USA}
\affiliation{Institute of Particle Physics: McGill University, Montr\'{e}al, Qu\'{e}bec, Canada H3A~2T8; Simon Fraser University, Burnaby, British Columbia, Canada V5A~1S6; University of Toronto, Toronto, Ontario, Canada M5S~1A7; and TRIUMF, Vancouver, British Columbia, Canada V6T~2A3}
\affiliation{University of Michigan, Ann Arbor, Michigan 48109, USA}
\affiliation{Michigan State University, East Lansing, Michigan 48824, USA}
\affiliation{Institution for Theoretical and Experimental Physics, ITEP, Moscow 117259, Russia}
\affiliation{University of New Mexico, Albuquerque, New Mexico 87131, USA}
\affiliation{The Ohio State University, Columbus, Ohio 43210, USA}
\affiliation{Okayama University, Okayama 700-8530, Japan}
\affiliation{Osaka City University, Osaka 588, Japan}
\affiliation{University of Oxford, Oxford OX1 3RH, United Kingdom}
\affiliation{Istituto Nazionale di Fisica Nucleare, Sezione di Padova-Trento, $^{ff}$University of Padova, I-35131 Padova, Italy}
\affiliation{University of Pennsylvania, Philadelphia, Pennsylvania 19104, USA}
\affiliation{Istituto Nazionale di Fisica Nucleare Pisa, $^{gg}$University of Pisa, $^{hh}$University of Siena and $^{ii}$Scuola Normale Superiore, I-56127 Pisa, Italy}
\affiliation{University of Pittsburgh, Pittsburgh, Pennsylvania 15260, USA}
\affiliation{Purdue University, West Lafayette, Indiana 47907, USA}
\affiliation{University of Rochester, Rochester, New York 14627, USA}
\affiliation{The Rockefeller University, New York, New York 10065, USA}
\affiliation{Istituto Nazionale di Fisica Nucleare, Sezione di Roma 1, $^{jj}$Sapienza Universit\`{a} di Roma, I-00185 Roma, Italy}
\affiliation{Rutgers University, Piscataway, New Jersey 08855, USA}
\affiliation{Texas A\&M University, College Station, Texas 77843, USA}
\affiliation{Istituto Nazionale di Fisica Nucleare Trieste/Udine, I-34100 Trieste, $^{kk}$University of Udine, I-33100 Udine, Italy}
\affiliation{University of Tsukuba, Tsukuba, Ibaraki 305, Japan}
\affiliation{Tufts University, Medford, Massachusetts 02155, USA}
\affiliation{University of Virginia, Charlottesville, Virginia 22906, USA}
\affiliation{Waseda University, Tokyo 169, Japan}
\affiliation{Wayne State University, Detroit, Michigan 48201, USA}
\affiliation{University of Wisconsin, Madison, Wisconsin 53706, USA}
\affiliation{Yale University, New Haven, Connecticut 06520, USA}

\author{T.~Aaltonen}
\affiliation{Division of High Energy Physics, Department of Physics, University of Helsinki and Helsinki Institute of Physics, FIN-00014, Helsinki, Finland}
\author{B.~\'{A}lvarez~Gonz\'{a}lez$^z$}
\affiliation{Instituto de Fisica de Cantabria, CSIC-University of Cantabria, 39005 Santander, Spain}
\author{S.~Amerio}
\affiliation{Istituto Nazionale di Fisica Nucleare, Sezione di Padova-Trento, $^{ff}$University of Padova, I-35131 Padova, Italy}
\author{D.~Amidei}
\affiliation{University of Michigan, Ann Arbor, Michigan 48109, USA}
\author{A.~Anastassov$^x$}
\affiliation{Fermi National Accelerator Laboratory, Batavia, Illinois 60510, USA}
\author{A.~Annovi}
\affiliation{Laboratori Nazionali di Frascati, Istituto Nazionale di Fisica Nucleare, I-00044 Frascati, Italy}
\author{J.~Antos}
\affiliation{Comenius University, 842 48 Bratislava, Slovakia; Institute of Experimental Physics, 040 01 Kosice, Slovakia}
\author{G.~Apollinari}
\affiliation{Fermi National Accelerator Laboratory, Batavia, Illinois 60510, USA}
\author{J.A.~Appel}
\affiliation{Fermi National Accelerator Laboratory, Batavia, Illinois 60510, USA}
\author{T.~Arisawa}
\affiliation{Waseda University, Tokyo 169, Japan}
% ...

\collaboration{CDF Collaboration\footnote{With visitors from
$^a$Istituto Nazionale di Fisica Nucleare, Sezione di Cagliari, 09042 Monserrato (Cagliari), Italy,
$^b$University of CA Irvine, Irvine, CA 92697, USA,
$^c$University of CA Santa Barbara, Santa Barbara, CA 93106, USA,
$^d$University of CA Santa Cruz, Santa Cruz, CA 95064, USA,
$^e$Institute of Physics, Academy of Sciences of the Czech Republic, Czech Republic,
$^f$CERN, CH-1211 Geneva, Switzerland,
$^g$Cornell University, Ithaca, NY 14853, USA,
$^h$University of Cyprus, Nicosia CY-1678, Cyprus,
$^i$Office of Science, U.S. Department of Energy, Washington, DC 20585, USA,
$^j$University College Dublin, Dublin 4, Ireland,
$^k$ETH, 8092 Zurich, Switzerland,
$^l$University of Fukui, Fukui City, Fukui Prefecture, Japan 910-0017,
$^m$Universidad Iberoamericana, Mexico D.F., Mexico,
$^n$University of Iowa, Iowa City, IA 52242, USA,
$^o$Kinki University, Higashi-Osaka City, Japan 577-8502,
$^p$Kansas State University, Manhattan, KS 66506, USA,
$^q$Ewha Womans University, Seoul, 120-750, Korea, %Korea University, Seoul, 136-713, Korea,
$^r$University of Manchester, Manchester M13 9PL, United Kingdom,
$^s$Queen Mary, University of London, London, E1 4NS, United Kingdom,
$^t$University of Melbourne, Victoria 3010, Australia,
$^u$Muons, Inc., Batavia, IL 60510, USA,
$^v$Nagasaki Institute of Applied Science, Nagasaki, Japan,
$^w$National Research Nuclear University, Moscow, Russia,
$^x$Northwestern University, Evanston, IL 60208, USA,
$^y$University of Notre Dame, Notre Dame, IN 46556, USA,
$^z$Universidad de Oviedo, E-33007 Oviedo, Spain,
$^{aa}$CNRS-IN2P3, Paris, F-75205 France,
$^{bb}$Texas Tech University, Lubbock, TX 79609, USA,
$^{cc}$Universidad Tecnica Federico Santa Maria, 110v Valparaiso, Chile,
$^{dd}$Yarmouk University, Irbid 211-63, Jordan,
}}
\noaffiliation

\date{\today}

\begin{abstract}
We present a search for the standard model Higgs boson produced in association with a $W^{\pm}$ boson.
This search uses data corresponding to an integrated luminosity of 7.5 fb$^{-1}$ collected by the CDF 
detector at the Tevatron. We select $WH \to \ell\nu b \bar{b}$ candidate events with two jets, large missing transverse energy, 
and exactly one charged lepton. We further require that at least 
one jet be identified to originate from a bottom quark. Discrimination between the signal and the
large background is achieved through the use of a Bayesian artificial neural network. The number of tagged events and their
distributions are consistent with the standard model expectations. 
We observe no evidence for a Higgs boson signal and set 95\% C.L. upper limits on the $WH$ production 
cross section times the 
branching ratio to decay to $b\bar b$ pairs, $\sigma(p\bar p \rightarrow W^{\pm} H) \times {\cal B}(H\rightarrow b\bar b)$, relative to the rate predicted by the standard model. For
the Higgs boson mass range of 100 GeV/c$^2$ to 150 GeV/c$^2$ we set observed (expected) upper limits from 
1.34 (1.83) to 38.8 (23.4). For 115 GeV/c$^2$ the upper limit is 
3.64 (2.78). The combination of the present search with an independent analysis that selects events with three jets yields more 
stringent limits ranging from 1.12 (1.79) to 34.4 (21.6) in the same mass range. 
For 115 and 125 GeV/c$^2$ the upper limits are 2.65 (2.60) and 4.36 (3.69), respectively.   

\end{abstract}

%  13.85.Rm  Limits on production of particles
%  14.80.Bn  Standard-model Higgs bosons

\maketitle

%%%%%%%%%%%%%%%%%%%%%%%%%%%%%%%%%%%%%%%%%%%%%%%%%%
% Section 1 Introduction %
%%%%%%%%%%%%%%%%%%%%%%%%%%%%%%%%%%%%%%%%%%%%%%%%%%
%% Currently, include from separated .tex file,
%% but will put all the text here once PRD is ready to submit.
%\input{introduction}
\section{Introduction}
The standard model (SM) describes not only the fundamental particles of 
quarks and leptons and their interactions, but also predicts the existence of a single 
scalar particle, the Higgs boson, which arises as a result of spontaneous electroweak 
symmetry breaking~\cite{Higgs:1964ia,Higgs:1964pj,Higgs:1964ghk,Higgs:1964eb}. 
The Higgs boson remains the only fundamental SM particle 
that has not been observed by experiment. Direct searches at LEP2~\cite{Barate:2003sz}, the 
Tevatron~\cite{TeVcombination11}, and recently LHC experiments~\cite{atlas,cms} have
constrained the Higgs boson mass to lie in the range between 115.5 and 127 GeV/c$^2$ at 95\% C.L.,
%have yielded constraints on the Higgs boson mass between 115.5 and 127 GeV/c$^2$ at 95\% C.L.,   
which is consistent with the 95\% C.L. upper limit of 152 GeV/c$^2$ obtained from global fits 
to precision electroweak data~\cite{EWGfits2009Summer}.    

In $\sqrt{s}=1.96\,\mathrm{TeV}$ proton-antiproton collisions, the Higgs boson is expected to be produced mainly 
through gluon fusion ($gg\rightarrow H$) and in association with a $W$ or $Z$ boson~\cite{Tev4LHC}. 
The cross section for $WH$ production is twice that of $ZH$ and is about a factor of 10 smaller than $gg\rightarrow H$.  
The Higgs boson decay branching fraction is dominated by $H \rightarrow b\bar{b}$ for the Higgs boson mass $m_H<135$ GeV/c$^2$ and
by $H \rightarrow W^+ W^-$ for $m_H>135$ GeV/c$^2$~\cite{Djouadi:1997yw}. 
A search for a low-mass Higgs ($m_H<135$ GeV/c$^2$) in the $gg\rightarrow H \rightarrow b\bar b$ channel is extremely challenging 
because the $b\bar b$ QCD production rate is many orders of magnitude larger than the Higgs boson production rate. 
Requiring the leptonic decay of the associated 
$W$ boson improves greatly the expected signal over background ratio in this 
channel. As a result, $WH\rightarrow l\nu b\bar b$~\cite{leptonNote} 
is one of the most promising channels for the low-mass Higgs boson searches, and it 
significantly contributes to the combined search for the Higgs boson at the Tevatron~\cite{TeVcombination11}. 

This paper presents a search for Higgs boson production in
proton-antiproton collisions using the $WH \rightarrow \ell \nu b\bar{b}$ 
channel at $\sqrt{s}=1.96\,\tev$ using data collected between February 2002 and March 2011 with the CDF detector.
The acquired data correspond to an integrated luminosity of approximately 7.5~$\invfb$.
Searches for the standard model Higgs boson using the same final state have been reported before 
by CDF~\cite{jason,me} and D0~\cite{d0} with data corresponding to an integrated luminosity of 
5.6~$\invfb$ and 5.3~$\invfb$, respectively.
Compared to the previously reported analysis, 
we have employed a Bayesian artificial neural network (BNN) discriminant~\cite{BNN,BNNbook}
to improve discrimination between signal and background. 
The signal acceptance is improved by using additional triggers based on jets and missing transverse energy, as well as a novel
method to combine them into a single analysis stream in order to maximize the event yield while properly accounting for correlations
between triggers. The signal acceptance is also increased by using several different lepton reconstruction algorithms, for muon and 
electron candidates. We have optimized $b$-tagging algorithms used in the analysis to increase signal acceptance. 
%However, there might be still room for further improvement. 
We also employed multivariate 
methods to improve the rejection of multi-jet QCD background, as well as to improve di-jet invariant mass resolution. 

Recently, the experiments at the Large Hadron Collider (LHC) have obtained enough data to set limits on the Higgs boson mass   
exceeding the sensitivity of the Tevatron experiments~\cite{atlas,cms}. 
However, at the LHC the most sensitive low-mass search is in the 
diphoton final state and searches for $H\rightarrow b\bar b$ will take more data before the Tevatron 
sensitivity is reached in this channel. In this sense, the Tevatron and LHC are complementary and both will provide
important information in the search for a low-mass Higgs boson. 

This paper is organized as follows.
Section~\ref{sec:apparatus} describes the experimental apparatus, 
the Collider Detector at Fermilab (CDF).
%In section~\ref{sec:b-tagging}, the method to identify $b$-jets is discussed.
Section~\ref{sec:selection} presents the data samples and the event selection used to identify the $WH \to \ell \nu b
\bar{b}$ candidate events.
Section~\ref{sec:bkg} presents the background modeling and its estimation. 
Section~\ref{sec:Acceptance} discusses the signal acceptance and its systematic
uncertainty.
Section~\ref{sec:Optimization} introduces advanced techniques to improve the
analysis sensitivity further.
The final results and conclusions are presented in Sec.~\ref{sec:results} and Sec.~\ref{sec:conclusions}.

%%%%%%%%%%%%%%%%%%%%%%%%%%%%%%%%%%%%%%%%%%%%%%%%%%
% Section 2 The CDF II detector %
%%%%%%%%%%%%%%%%%%%%%%%%%%%%%%%%%%%%%%%%%%%%%%%%%%
%\input{detector}
%\section{Experimental Apparatus}
\section{The CDF II Detector}
\label{sec:apparatus}

The CDF II detector~\cite{Acosta:2004yw} geometry is described using a 
cylindrical coordinate system.  The $z$-axis follows the
proton direction, and the polar angle $\theta$ is usually expressed
through the pseudorapidity $\eta = -\ln(\tan(\theta/2))$.  The
detector is approximately symmetric around $\eta=0$ and in the azimuthal
angle~$\phi$.  The energy transverse to the beam is defined as $E_T=E\sin \theta$,
and the momentum transverse to the beam is $p_T=p \sin \theta$.

Charged particles are tracked by a system of silicon microstrip
detectors~\cite{Sill:2000svx} and a large open cell drift 
chamber~\cite{Affolder:2004cot} in the region
$|\eta|\leq 2.0$ and $|\eta|\leq 1.0$, respectively.  The tracking
detectors are immersed in a $1.4\,\mathrm{T}$ solenoidal magnetic
field aligned with the incoming beams, allowing measurement
of charged particle $p_T$.

The transverse momentum resolution is measured to be $\delta p_T/p_T
\approx 0.07\% \cdot p_T$(GeV/c) for the combined tracking 
system~\cite{Acosta:2004yw}.
The resolution on the track impact parameter ($d_0$), the distance from
the beam-line axis to the track at the track's closest approach in the
transverse plane, is $\sigma(d_0) \approx 40\,\mu{\rm m}$, of which about
$30\,\mu{\rm m}$ is due to the transverse size of the
Tevatron beam itself~\cite{Sill:2000svx}.

Outside of the tracking systems and the solenoid, segmented
calorimeters with projective tower geometry are used to reconstruct
electromagnetic showers and hadronic
jets~\cite{Balka:1987ty,Bertolucci:1987zn,Albrow:2001jw} over the
pseudorapidity range $|\eta|<3.6$.  The transverse energy is measured 
in each calorimeter tower where the polar
angle ($\theta$) is calculated using the measured $z$ position of the event
vertex and the tower location.

Contiguous groups of calorimeter towers with signals are
identified and summed together into an energy cluster.  Electron
candidates are identified in the central electromagnetic calorimeter
(CEM) or in the forward, known as the plug, electromagnetic calorimeter (PEM) 
as isolated, mostly electromagnetic, clusters that match a reconstructed silicon track
in the pseudorapidity range $|\eta|<1.1$ and $1.1<|\eta| < 2.0$, respectively.  
The electron transverse
energy is reconstructed from the electromagnetic cluster with a
precision $\sigma(E_T)/E_T \approx 13.5\%/\sqrt{E_T(\mathrm{GeV})} \oplus
2\%$ for central electrons~\cite{Balka:1987ty} and 
$\sigma(E_T)/E_T = 16.0\%/\sqrt{E_T(\mathrm{GeV})} \oplus
2\%$ for plug electrons~\cite{Breccia:2004}.  Jets are identified as a group of
electromagnetic calorimeter energy ($E_{EM}$) and hadronic calorimeter energy ($E_{HAD}$) 
clusters populating a cone of radius $\Delta{R} \approx\sqrt{(\Delta
\phi)^2 + (\Delta \eta)^2} \leq 0.4$ units around a high-$E_T$ seed
cluster~\cite{Abe:1991ui}.  Jet energies are corrected for calorimeter
nonlinearity, losses in the gaps between towers, and
multiple primary interactions. The jet energy resolution is
approximately $\sigma(E_T) \approx \left[0.1 E_T + 1.0~{\rm GeV}\right]$~\cite{Bhatti:2005ai}.

Muon candidates are detected in three separate subdetectors.
After at least five interaction lengths in the calorimeter, central muons first
encounter four layers of planar drift chambers (CMU), capable of
detecting muons with $p_T > 1.4\,{\rm GeV}/c$~\cite{Ascoli:1987av}.  Four
additional layers of planar drift chambers (CMP) behind another 60~cm of steel detect muons
with $p_T > 2.8$ GeV/$c$~\cite{Dorigo:2000ip}.  These two systems cover the same
central pseudorapidity region with $|\eta| \leq 0.6$. A track that is linked to
both CMU and CMP stubs is called a CMUP muon. Muons that
exit the calorimeters at $ 0.6 \leq |\eta| \leq 1.0$ are detected by
the CMX system of four drift layers.  Muon candidates are then
identified as isolated tracks that extrapolate to line segments or
``stubs'' in the muon subdetectors. 

Missing transverse energy ({\MET}) is 
defined as the opposite of the vector sum
of all calorimeter tower energy depositions projected on the transverse
plane.  It is used as a measure of the sum
of the transverse momenta of the particles that escape detection, most
notably neutrinos.  
The corrected energies are used for jets in the vector sum defining {\MET}.
The muon momentum is 
also added for any minimum ionizing high-$p_T$ muon found in the event.

Muon and electron candidates used in this analysis are identified during data taking 
with the CDF trigger 
system, a three-level filter with tracking information available at the first 
level~\cite{Thomson:2002xp}.
%Events used in this analysis have all passed the high-energy electron
%or muon trigger selection.  
The first stage of the central electron
trigger (CEM) requires a track with $p_T > 8$~GeV/$c$ pointing to a tower with
$E_T > 8$~GeV and $E_{\mathrm{HAD}}/E_{\mathrm{EM}}<0.125$. As appropriate for 
selecting $W$-decay electrons, the plug electron trigger (MET+PEM) requires a 
tower with $E_T > 8$~GeV, $E_{\mathrm{HAD}}/E_{\mathrm{EM}}<0.125$ and the missing 
transverse energy {\MET} $> 15$ GeV.  The first
stage of the muon trigger requires a track with $p_T > 4$~GeV/$c$
(CMUP) or 8~GeV/$c$ (CMX) pointing to a muon stub.  A complete lepton
reconstruction is performed online in the final trigger stage, where
we require $E_T > 18\,\mathrm{GeV}$ for central electrons (CEM), 
$E_T > 18\,\mathrm{GeV}$ and {\MET}$> 20\, \mathrm{GeV}$ for plug 
electron (MET+PEM) and $p_T > 18\,\mathrm{GeV}/c$ for muons (CMUP, CMX).

The {\MET} + 2 jet trigger has been previously used in the $WH$ analysis~\cite{jason}, which 
complements the high-$p_T$ lepton triggers by identifying a lepton from $WH$ decay as a 
high-$p_T$ track isolated from other tracks which has failed the 
lepton triggers mentioned above. At high instantaneous Tevatron luminosity,
the accept rate of this trigger is reduced (pre-scaled) by randomly sampling a luminosity-dependent
fraction of events.
This trigger also requires two jets with $E_T>10$~GeV, one of them central ($|\eta|<1.1$), 
and {\MET}$>35$~GeV. We also include a second {\MET} and two-jets trigger,
which was introduced only in the second part of the data and 
requires two jets with $E_T>$10~GeV and {\MET}$>30$~GeV.
We also include a third trigger based on {\MET} only, and {\MET}$>$45~GeV 
for the first part of the data, while the selection criteria is relaxed to 40 GeV for the second part of 
the data.
 
The efficiency of the different triggers is
measured using the lepton triggered data and is parametrized using sigmoid turn-on curves 
as a function of {\MET}, without correcting for the muon momenta.
The novel method exploited to combine and parametrize all the three trigger paths is described in~\cite{BuzatuThesis},
which can be generalized to any combination of different trigger paths, allowing optimal performance.    
    
%%%%%%%%%%%%%%%%%%%%%%%%%%%%%%%%%%%%%%%%%%%%%%%%%%
% Section 4 Event Selection %
%%%%%%%%%%%%%%%%%%%%%%%%%%%%%%%%%%%%%%%%%%%%%%%%%%
%\input{selection}
\section{Data Samples and Event Selection}
\label{sec:selection}

%The results presented here use data collected between February 2002 and March 2011. 
The data collected using the lepton-based (CEM, CMUP, CMX and MET+PEM)
triggers correspond to $7.5 \pm 0.4$~fb$^{-1}$ of integrated luminosity, while the data from the
{\MET}-based triggers correspond to $7.3 \pm 0.4$~fb$^{-1}$.

The $WH\rightarrow \ell\nu b\bar b$
signal consists of two $b$ jets, a high-$p_T$ lepton, and large missing energy. This section provides an 
overview of the signal reconstruction with a focus on the improvements of this analysis over a previous $WH$ 
search~\cite{jason}.

\subsection{Improving Lepton Identification} 
We use several different lepton identification algorithms in order to include events from 
multiple trigger paths. Each algorithm requires a single high-$p_T$ ($>20$ GeV/c), isolated charged 
lepton consistent with leptonic $W$ boson decay. Because the lepton from a leptonic $W$ decay 
is well-isolated from the rest of the event, the additional energy in 
the cone of $\Delta R=0.4$ surrounding the lepton
is required to have less than 10\% of the lepton energy. We employ the same lepton identification
algorithms as the prior CDF $WH$ search~\cite{jason}. The tight lepton is required to be identified as
either an electron (CEM, PEM), a muon (CMUP, CMX), or an isolated track from the data collected with
{\MET} triggers. 

We further improve the lepton acceptance by about 10\% by including two
additional lepton identification algorithms. One lepton type is selected
from CEM-triggered events using a multivariate likelihood method to select
electron candidates that fail the standard electron requirements. Another
lepton type is selected from {\MET}-triggered events by requiring an
isolated track with significant deposits of energy in the calorimeter.
Such tracks primarily originate from the leptonic decay of the $W$ boson,
where the electrons fail the standard identification, or from $\tau$
leptons that decay into single charged hadrons.

The efficiency of lepton identification is measured using $Z\rightarrow e^+e^-$ and $Z\rightarrow \mu^+\mu^-$
samples. A pure sample of leptons is obtained by selecting events where the invariant mass of two high-$p_T$
tracks is near the mass of the $Z$ boson and one track passed the trigger and tight lepton selection. 
The efficiency is then measured using the other unbiased track. 
The same procedure is applied to simulated events events and a correction factor is applied to correct
the difference due to imperfect detector modeling.
 
\subsection{$b$-jet Identification}

Multijet final states have dominant contributions from QCD light-flavor jet production. The low-mass standard model
Higgs boson decays predominantly to $b$-quark pairs. Jets from $b$ quarks can be distinguished from light-flavor
jets by looking for the decay of long-lived $B$ hadrons within the jet cone. 
We employ three $b$-identification algorithms to 
optimize the selection of $b$-quark jets. The secondary vertex tagging algorithm~\cite{Acosta:2004hw} 
(SECVTX) attempts to reconstruct a secondary vertex using tracks found within a jet. 
If a vertex is found and it is significantly displaced 
from the $p\bar p$ interaction point (primary vertex), the jet is identified as a $b$-jet(``$b$-tagged'').
The Jet Probability algorithm~\cite{Abulencia:2006kv} (JP) uses tracking information from tracks inside a jet 
to identify $B$ decays. The algorithm looks at the distribution of impact parameters 
for tracks inside a jet to form a probability that the jet originated from the primary vertex.
Light jets yield a probability distribution approximately constant between 0 and 1, while $b$ jets preferentially 
populate low values of probability.
A jet is considered as $b$-tagged if the jet probability
value is less than 5\%. The neural network tagging algorithm~\cite{roma} (NN) combines the strengths 
of existing $b$-tagging information more efficiently using a multivariate technique exploiting variables 
such as displaced vertices, displaced tracks, and low-$p_T$ muons from $b$-quark decay. 
The NN provides an output value ranging from -1~(light-jet-like) to 1~($b$-jet-like). The cut on 
this continuous output has been tuned to provide maximum sensitivity: a jet is considered as 
$b$-tagged if the jet's NN output is positive ($>0$).

To increase the signal to background ratio for $WH$ events, at least one jet must be $b$-tagged by the SECVTX 
algorithm. We then divide our sample into four exclusive categories in a preferential order 
based on the purity of $b$-tagged jets. The first category (ST+ST) comprises events 
where there are two SECVTX $b$-tagged jets. The second category (ST+JP) consists of events where only one of the 
jets is $b$-tagged by SECVTX and the second jet is $b$-tagged only by JP. The third category (ST+NN)
is similar to the second, but the second jet is $b$-tagged only by NN. The fourth category (ST) contains events 
where only one of the jets is $b$-tagged by SECVTX and the second jet is not $b$-tagged. 

\subsection{Lepton + Jets Selection} 

After identifying the final state in the event, we require that the events 
contain 
%one electron with $E_T>20$ GeV or muon with $p_T>20$ GeV/c, 
one high-$p_T$ lepton ($>$ 20 GeV/c), 
corrected {\MET}$>20$ GeV (25 GeV in the case of forward electrons), and two jets with 
corrected $E_T>20$ GeV and $|\eta|<2.0$.  The event's primary vertex is calculated by fitting a subset of 
well-measured tracks coming from the beam line and is required to be within 60 cm of 
the center of the CDF II detector~\cite{Acosta:2004yw}.
The longitudinal coordinate $z_0$ of the lepton track at point of closest approach to the beam line must 
be within 5 cm of the primary vertex to ensure that the lepton and the jets come from the same hard interaction. 
In order to reduce the $Z$ + jets and $WW/WZ$ background rates, events with more than one lepton are rejected. 
Events from $Z\rightarrow l^+l^-$ decays in which one lepton is 
not identified are removed by vetoing events where the invariant mass of the lepton and 
any track in the event is within the $Z$ mass window between 76 and 106 GeV/c$^2$.

Before applying any $b$-tagging algorithm, the sample (pretag sample) has dominant contributions from 
$W$ + jets and QCD multijet production.
We use the $b$-tagging strategies outlined above to increase the signal purity of the $W$ + 2 jet events. 
We further purify the sample with exactly one secondary vertex tagged jet (ST) by applying additional 
kinematic and angular cuts to reduce QCD multijet events that mimic the $W$-boson signature. 
The rejection is based on a support vector machine multivariate discriminant that was 
optimized to identify the $W$ + jets events against the QCD events~\cite{svm}.

%%%%%%%%%%%%%%%%%%%%%%%%%%%%%%%%%%%%%%%%%%%%%%%%%%
% Section 3 Identification of b-jet %
%%%%%%%%%%%%%%%%%%%%%%%%%%%%%%%%%%%%%%%%%%%%%%%%%%
%\input{btagging}

%%%%%%%%%%%%%%%%%%%%%%%%%%%%%%%%%%%%%%%%%%%%%%%%%%
% Section 5 Background Estimation %
%%%%%%%%%%%%%%%%%%%%%%%%%%%%%%%%%%%%%%%%%%%%%%%%%%
%\input{background}
\section{Background Estimation}
\label{sec:bkg}

The final state signature of $WH\rightarrow l\nu b\bar b$ production can be mimicked by a number of 
processes. The dominant backgrounds are $W$ + jets production, $t\bar t$ production, single top
production, and QCD multijet production. Several electroweak production processes 
(diboson or $Z$ + jets) also contribute with smaller rates. We estimate the background rates 
based on the same strategies used 
in the previous top cross section measurement~\cite{Acosta:2004hw}, 
single top searches~\cite{singletop}, and $WH$ analysis~\cite{jason}. We provide an overview of each 
background estimate below.

\subsection{Top and Electroweak Backgrounds} 
\label{sec:topbkg}
Production of both top-quark pairs and single top quarks contributes to
the tagged $W$+jets sample. Several electroweak boson production
processes also contribute. Pairs of $WW$ can decay to a lepton, a neutrino
(seen as missing energy), and two jets, one of which may originate from a charm quark. Pairs of $WZ$
events can decay to the signal $\ell \nu b\bar b$ or $\ell \nu c\bar c$ final
state.  Finally, $Z\rightarrow\tau^+\tau^-$ events with one
leptonic $\tau$ decay and one hadronic decay contribute, yielding a lepton, missing traverse energy, and a 
narrow jet displaced from the primary interaction point.

The normalizations of the diboson and top production backgrounds are based
on the theoretical cross sections~\cite{Campbell:2002tg,Cacciari:2003fi,Harris:2002md} 
listed in Table~\ref{tbl:xsec}, the time-integrated
luminosity, and the acceptance and $b$-tagging efficiency derived from Monte Carlo
events. The acceptance is corrected based on measurements using data for lepton identification, trigger
efficiencies, $b$-tagging efficiencies, and the $z$ vertex cut. 
%The tagging efficiency is corrected by the $b$-tagging scale factor.
The total top and electroweak contributions in each tagging 
category are shown in Table~\ref{Table:BGsumarry}. We use the measured inclusive cross 
section (787.4$\pm$85.0 pb) for $Z$ + jets~\cite{Acosta:2004uq}.

\begin{table}
  \begin{center}
    \caption{Theoretical
    cross sections and uncertainties for the electroweak and single top
    backgrounds, along with the theoretical cross section for
    $t\bar{t}$ at $m_t = 172.5\,\mathrm{GeV}/c^2$.}
    \begin{tabular}{cc}
      \hline \hline
      Background & Theoretical cross sections [pb]\\ \hline  
	$WW$ & 11.66 $\pm$ 0.70 \\ 
	$WZ$ & 3.46 $\pm$ 0.30 \\ 
	$ZZ$ & 1.51 $\pm$ 0.20 \\ 
	single-top $s$-channel & 1.05 $\pm$ 0.07 \\ 
	single-top $t$-channel & 2.10 $\pm$ 0.19 \\
	$t\bar{t}$ & 7.04 $\pm$ 0.44 \\
    \hline\hline
    \end{tabular}
    \label{tbl:xsec}
  \end{center}
\end{table}

\subsection{$W$ + heavy flavor}
\label{sec:hfbkg} 
The $Wb\bar{b}$, $Wc\bar{c}$, and $Wc$ processes ($W$ + heavy flavor) are major background
sources after the $b$-tagging requirement. 
Large theoretical uncertainties exist for the overall
normalization because current Monte Carlo
event generators can generate $W$+heavy-flavor events only to tree-level. 
Consequently, the rates for these processes are normalized to data. 
The contribution from true heavy-flavor production in
$W$+jets events is determined from measurements of the heavy-flavor
event fraction in $W+$jets events and the $b$-tagging efficiency for
those events.

The fraction of $W$+jets events produced with heavy-flavor jets has
been studied extensively using a combination of {\sc alpgen}~+~{\sc pythia}
Monte Carlo generators~\cite{Mangano:2002ea,Corcella:2001wc,Sjostrand:2000wi}.  Calculations of the
heavy-flavor fraction in {\sc alpgen} have been calibrated using a jet 
data sample, and a scaling factor of $1.4\pm0.4$
is necessary to make the heavy-flavor production in Monte Carlo match
the production in $W$+1 jet events. 

For the tagged $W$+heavy flavor (HF) background estimate, the heavy-flavor fractions
and tagging rates are multiplied by the number of pretag
$W$+jets candidate events ($N_{\rm pretag}$) in data, after correction for the
contribution of non-$W$ ($f_{{\rm non}-W}$) as determined from the fits described in Section~\ref{sec:nonwbkg}
, $t\bar{t}$, and other background events to the pretag sample.
The $W$+heavy flavor background contribution is obtained by the following
relation:
%
%\begin{widetext}
\begin{equation}
  N_{W+{\rm HF}} = f_{{\rm HF}}  \epsilon _{{\rm tag}} 
  \left [ N_{\rm pretag} (1-f_{{\rm non}-W}) - N_{{\rm TOP}} - N_{{\rm
  EWK}}\right ],
\end{equation}
%\end{widetext}
where $f_{HF}$ is the heavy-flavor fraction, $\epsilon_{\rm tag}$ is
the tagging efficiency, 
$N_{\rm TOP}$ is the expected number of $t\bar
t$ and single top events, and $N_{\rm EWK}$ is the expected background contribution from
$WW$, $WZ$, $ZZ$ and $Z$ boson events, as described in Section~\ref{sec:topbkg}.

The total $W$ + heavy flavor contributions in each tagging 
category are shown in Table~\ref{Table:BGsumarry}. 

%%%%%%%%%%%%%%% All lepton types merged tables %%%%%%%%%%%%%%%%%%%%%
\begin{table*}
  \begin{center}
  \caption{Background summary table for each $b$-tagging category after all lepton categories combined. 
    As a reference, the expected signal for $m_H = 115~\gevcc$ is also shown.
  }
  \begin{tabular}{ccccc}\hline \hline
                        &    ST+ST          &          ST+JP  &     ST+NN   & 1-ST \\ \hline
          Pretag events &                                   \multicolumn{4}{c}{184050} \\ \hline
      $t\bar{t}$        &   142  $\pm$ 22   & 114  $\pm$ 12   &  62.8 $\pm$ 6.4  & 479  $\pm$ 49\\
       Single top ($s$-ch) &   45.0 $\pm$ 6.7  & 35.1 $\pm$ 3.4  &  18.9 $\pm$ 1.8  & 106  $\pm$ 10\\
       Single top ($t$-ch) &   13.9 $\pm$ 2.4  & 13.3 $\pm$ 2.0  &  8.7  $\pm$ 1.2  & 191  $\pm$ 23\\
                   $WW$ &   1.67 $\pm$ 0.42 & 6.23 $\pm$ 2.08 &  5.14 $\pm$ 1.35 & 186  $\pm$ 25\\
                   $WZ$ &   12.9 $\pm$ 2.0  & 10.7 $\pm$ 1.2  &  5.84 $\pm$ 0.62 & 53.3 $\pm$ 6.2\\
                   $ZZ$ &   0.62 $\pm$ 0.09 & 0.49 $\pm$ 0.06 &  0.29 $\pm$ 0.03 & 2.05 $\pm$ 0.23\\
               $Z$ + jets &   9.64 $\pm$ 1.40 & 11.9 $\pm$ 1.7  &  8.75 $\pm$ 1.30 & 182  $\pm$ 25\\
            $Wb\bar{b}$ &   257  $\pm$ 104  & 228  $\pm$ 91   &  125  $\pm$ 50   & 1450 $\pm$ 580\\
          $Wc\bar{c}/c$ &   31.0 $\pm$ 12.6 & 98.3 $\pm$ 40.5 &  63.8 $\pm$ 26.0 & 1761 $\pm$ 708\\
                 Mistag &   12.1 $\pm$ 2.9  & 52.8 $\pm$ 15.2 &  57.0 $\pm$ 14.3 & 1646 $\pm$ 220\\
            Non-$W$ QCD &   57.9 $\pm$ 23.6 & 85.3 $\pm$ 34.1 &  74.9 $\pm$ 29.9 & 747  $\pm$ 299\\
 \hline
       Total background &   584  $\pm$ 169  & 656  $\pm$ 194  &  432  $\pm$ 126  & 6802 $\pm$ 1822\\ \hline
       Observed events &   519  &     568            &     402        &  6482  \\ \hline
%     $WH$ (115~GeV) &                3.21 &Control region &  Control region \\ \hline
%     $ZH$ (115~GeV) &                3.21 &Control region &  Control region \\ \hline
$WH$ and $ZH$ signal (115~GeV/c$^2$) & 7.28 & 5.34 & 2.80 & 16.0 \\ \hline
%%$WH$ and $ZH$ signal (115~GeV) & 7.28 $\pm$ 0.98 & 5.34 $\pm$ 0.39 & 2.80 $\pm$ 0.19 & 16.0 $\pm$ 1.2 \\ \hline
 \end{tabular}
  \label{Table:BGsumarry}
  \end{center}
\end{table*}

\subsection{Non-$W$ QCD Multijet}
\label{sec:nonwbkg} 

Events from QCD multijet production may mimic the
$W$-boson signature due to instrumental background.  
When a jet passes the charged lepton selection criteria
or a heavy-flavor jet produces a charged leptons via semileptonic decay, the jet
is reconstructed incorrectly as a charged lepton, which is denoted as a non-$W$ lepton. 
Non-$W$ {\MET} can result from mismeasurements of energy or
semileptonic decays of heavy-flavor quarks. Since the {\MET} mismeasurement
is usually not well modeled in the detector simulation, we use
several different samples of observed events to model the non-$W$ multijet contribution. 
One sample is based on events that fired the central electron trigger but failed at least two of the five
electron selection identification requirements that do not depend on the kinematic
properties of the event, such as the fraction of energy in the hadronic calorimeter. This sample is used
to estimate the non-$W$ contribution from CEM, CMUP, and CMX events. A second sample is formed
from events that pass a generic jet trigger with transverse energy $E_T>20$ GeV to model PEM events. 
These jets are additionally required to have a fraction of energy deposited in the 
electromagnetic calorimeter between 80\% and 95\%, and fewer than four tracks, to mimic electrons.
A third sample, used to model the non-$W$ background in isolated track events, consists of events 
that are required to pass the {\MET} triggers and contain a muon that passes all identification
requirements but fails the isolation requirement. 

To estimate the non-$W$ fraction in both the pretag and tagged sample, the {\MET} spectrum is fit to a sum of
the predicted background shapes. The fit has one fixed component and two templates whose normalization can 
float. The fixed component is obtained by adding the contributions of the simulated processes 
based on theoretical cross sections. The two floating
templates are a Monte Carlo $W$ + jets template and a non-$W$ template. The non-$W$ template is different
depending on the lepton category, as explained above. 
%The pretag non-$W$ fraction is used to estimate the heavy flavor and light flavor fractions. 
The total non-$W$ contribution for each tagging 
category is also shown in Table~\ref{Table:BGsumarry}. 

\subsection{Mistagged Jets} 
\label{sec:mistagbkg} 

Events with $W$ + light-flavor jets containing no $b$ or $c$ quark with a fake $b$ tag (mistags) 
can contribute to our tagged signal sample. 
We estimate the amount of mistags using the number of pretag $W$ + light flavor events and the
event mistag probability. The amount of pretag $W$ + light flavor is determined from the 
pretag sample by subtracting the events from non-$W$, top and electroweak, and $W$ + heavy flavor 
contributions. The event mistag probability is based on the per-jet mistag matrix that is derived from 
inclusive jet data by counting the number of false tags per jet for each $b$ tagger and 
is parametrized as a function of jet $E_T$, $\eta$, number of vertices, track multiplicity, and 
the scalar sum of jet $E_T$ in the event. For each event in our $W$ + light flavor Monte Carlo samples, 
we apply the per-jet mistag matrix to each jet and combine the probability to get an event mistag 
probability. The total mistag contribution for each tagging 
category is also shown in Table~\ref{Table:BGsumarry}.

\subsection{Summary of Background Estimation} 
\label{sec:sumbkg} 
 
The summary of the background  and signal ($m_H$ = 115 GeV/c$^2$) estimates and 
the number of observed events are shown in Table~\ref{Table:BGsumarry} for each 
tagging category. In this table, all lepton types are combined.
In general, the numbers of expected and observed events are in good agreement.

%%%%%%%%%%%%%%%%%%%%%%%%%%%%%%%%%%%%%%%%%%%%%%%%%%%%%%%%%%%%
% Section 6 Signal Acceptance and Systematic Uncertainties %
%%%%%%%%%%%%%%%%%%%%%%%%%%%%%%%%%%%%%%%%%%%%%%%%%%%%%%%%%%%%
%\input{signal}
\section{Signal Acceptance}
\label{sec:Acceptance}

In this section, the number of expected Higgs events and systematic uncertainties on the signal 
acceptance are discussed. We consider the signal acceptance for 
the $WH \to \ell \nu b \bar{b}$ process and the residual contribution of 
$ZH \to \ell \ell \!\!\!\!\!\!\not\,\,\,\, b \bar{b}$ where one of the leptons fails the $Z$ removal cut. 
We generated $WH\rightarrow l\nu b \bar{b}$ and $ZH\rightarrow l^+l^- b\bar b$ samples using the 
{\pythia} Monte Carlo program~\cite{Sjostrand:2000wi} for 11 values of the SM Higgs mass sampled between 100 and 150 GeV/c$^2$. 
The number of expected $WH\rightarrow l\nu b\bar b$ events ($N$) is given by 
\begin{equation} 
N = \epsilon {\cal{L}} 
%\cdot \sigma (p \bar{p} \rightarrow WH)\cdot \branchingratio(H \rightarrow b \bar{b}), \label{eq:ExpEvt}  
 \sigma (p \bar{p} \rightarrow WH) {\cal B}(H \rightarrow b \bar{b}), \label{eq:ExpEvt} 
\end{equation}
where $\epsilon$,  $\sigma(p \bar{p} \rightarrow WH)$, and 
%$\branchingratio(H \rightarrow b \bar{b})$ are the event detection efficiency, integrated luminosity,
${\cal B}(H \rightarrow b \bar{b})$ are the event detection efficiency,
production cross section, and branching ratio, respectively, and $\cal{L}$ is the integrated luminosity of the data-taking period.
The production cross section and 
branching ratio are calculated to next-to-leading order (NLO) precision~\cite{Tev4LHC}. 

The total event detection efficiency is the product of several efficiencies: the trigger efficiency, the primary vertex 
reconstruction efficiency, the lepton identification efficiency, the $b$-tagging
efficiency, and the event kinematic selection efficiency. 
The lepton trigger efficiency is measured using a clean $W\rightarrow l\nu$ data sample, 
obtained from other triggers after applying more stringent offline cuts. 
The {\MET} trigger efficiency is obtained using a trigger combination 
method~\cite{BuzatuThesis}.
The primary vertex efficiency is obtained using the vertex distribution from the minimum bias data. 
The lepton identification efficiency is calculated using 
$Z\rightarrow l^+l^-$ data and Monte Carlo samples. The $b$-tag efficiency is 
measured in a $b$-enriched sample from semileptonic heavy flavor decay.

The expected number of signal events is estimated for each of the probed values of the Higgs boson mass.  
Table~\ref{Table:BGsumarry} shows the number of expected $WH$ and $ZH$ events for $M_H=$ 115 GeV/c$^2$ in 
7.5 fb$^{-1}$. 

% \begin{table}
% \begin{center}
%   \begin{tabular}{cccccccc}
%     \hline  
%     \hline  
%      Category  & CEM & PHX & CMUP & CMX & {ISOTRK} & Loose Electron & All Leptons\\
%     \hline
%     ST+ST & 2.40 & 0.36 & 1.38 & 0.71 & 1.69 & 0.74 & 7.28\\
%     ST+JP & 1.73 & 0.25 & 1.03 & 0.53 & 1.25 & 0.55 & 5.34\\
%     ST+NN & 0.92 & 0.14 & 0.53 & 0.27 & 0.64 & 0.30 & 2.80\\
%      1-ST & 5.20 & 0.93 & 2.95 & 1.50 & 3.64 & 1.74 & 16.0\\
%     \hline 
%     All tags & 10.3 & 1.68 & 5.89 & 3.01 & 7.22 & 3.33 & 31.4\\
%     \hline 
%     \hline 
%   \end{tabular}
%   \caption{Expected number of $WH$ events for $m_{H}=$~115~$\gevcc$,
%   shown for each $b$-tag category and lepton type.
%   }
%   \label{table:expectedSignal}
% \end{center}
% \end{table}

The total systematic uncertainty on the acceptance comes from several sources, including trigger efficiencies, the 
jet energy scale, initial and final state radiation, lepton identification,  
luminosity, and $b$-tagging efficiencies. 
The lepton trigger uncertainties are measured using $Z$ boson decays.
The acceptance uncertainty due to the jet energy scale (JES)~\cite{Bhatti:2005ai} 
is calculated by shifting jet energies in $WH$
Monte Carlo samples by $\pm$ one standard deviation. The deviation from the nominal acceptance is taken as the 
systematic uncertainty. 
We estimate the impact of changes in initial state radiation (ISR) and final state radiation (FSR) by halving and doubling the 
parameters related to initial and final state radiation in the Monte Carlo event generation~\cite{Abulencia:2005aj}. 
The difference from the nominal acceptance is taken as the systematic uncertainty. 
The uncertainty in the incoming partons' energies relies on the parton distribution function (PDF) fits.
A NLO version of the PDFs, CTEQ6M, provides a 90\% confidence interval for each of the eigenvector input parameters~\cite{Pumplin:2002vw}. 
The nominal PDF value is reweighted to have a 90\% confidence level value, and the corresponding
reweighted acceptance is computed. The differences between the nominal and the reweighted 
acceptance are added in quadrature, and the total is assigned as the systematic uncertainty~\cite{Acosta:2004hw}.

The lepton identification uncertainties are estimated based on studies comparing $Z\rightarrow l^+l^-$ 
events in data and Monte Carlo. 

The systematic uncertainty of 6\% in the CDF luminosity 
measurement is treated as fully correlated between the signal and all Monte 
Carlo based background samples.
%For the signal sample and all Monte Carlo based samples a systematic uncertainty is applied for the 
%uncertainty of 6\% in the CDF luminosity measurement as fully correlated.
% which is common across all samples and channels. 
 
%The effect of the $b$-tagging scale factor uncertainty is determined from the background estimate.
The systematic uncertainty on the event tagging efficiency is estimated by varying the $b$-tagging efficiency 
and mistag prediction by $\pm$ one standard deviation and calculating the difference between the shifted acceptance and the 
default one. 
 
Total systematic uncertainties are summarized in
Tables~\ref{table:SigSystematic_central}, \ref{table:SigSystematic_plug}, and~\ref{table:SigSystematic_loose}.

\begin{table*} 
  \begin{center} 
\caption{Systematic uncertainties on the acceptance for central leptons (in percent).} 
\begin{tabular}{ccccccc}\hline \hline 
Category &JES & ISR/FSR/PDF & Lepton ID & Trigger & $b$-tag & Total\\ \hline 
ST+ST & 2.0 & 4.9 & 2 & $<$ 1 & 8.6  & 10.3  \\ 
ST+JP & 2.8 & 4.9 & 2 & $<$ 1 & 8.1  & 10.1  \\ 
ST+NN & 2.2 & 7.7 & 2 & $<$ 1 & 13.6 & 15.9  \\ 
1-ST  & 2.3 & 3.0 & 2 & $<$ 1 & 4.3  & 6.1   \\
\hline
\end{tabular} 
  \label{table:SigSystematic_central} 
  \end{center} 
\end{table*} 

\begin{table*} 
  \begin{center} 
\caption{Systematic uncertainties on the acceptance for forward electrons (in percent).} 
\begin{tabular}{ccccccc}\hline \hline 
Category & JES & ISR/FSR/PDF & Lepton ID & Trigger & $b$-tag & Total\\ \hline 
ST+ST & 2.4  & 7.7  & 2 & $<$ 1 & 8.6   & 12.0  \\ 
ST+JP & 3.9  & 4.5  & 2 & $<$ 1 & 8.1   & 10.3  \\ 
ST+NN & 6.7  & 12.9 & 2 & $<$ 1 & 13.6  & 20.0  \\ 
1-ST  & 2.9  & 5.7  & 2 & $<$ 1 & 4.3   & 8.0   \\
\hline
\end{tabular} 
  \label{table:SigSystematic_plug} 
  \end{center} 
\end{table*} 

\begin{table*} 
  \begin{center} 
\caption{Systematic uncertainties on the acceptance for additional 
leptons (in percent).} 
\begin{tabular}{ccccccc}\hline \hline 
Category & JES & ISR/FSR/PDF & Lepton ID & Trigger & $b$-tag & Total\\ \hline 
ST+ST & 1.7 & 7.1   & 4.5 & 3.0  & 8.6    & 12.5  \\ 
ST+JP & 2.4 & 6.4   & 4.5 & 3.0  & 8.1    & 11.9  \\ 
ST+NN & 1.9 & 19.5  & 4.5 & 3.0  & 13.6   & 24.5  \\ 
1-ST  & 4.7 & 8.4   & 4.5 & 3.0  & 4.3    & 11.8  \\
\hline
\end{tabular} 
  \label{table:SigSystematic_loose} 
  \end{center} 
\end{table*}

%%%%%%%%%%%%%%%%%%%%%%%%%%%%%%%%%%%%%%%%%%%%%%%%%%
% Section 7 Analysis Optimization %
%%%%%%%%%%%%%%%%%%%%%%%%%%%%%%%%%%%%%%%%%%%%%%%%%%
%\input{optimization}
\section{Analysis Optimization}
\label{sec:Optimization}
In this section we discuss the analysis optimization procedure after the event selection. 

\subsection{$b$-jet Energy Correction}
\label{sec:bjec}
The dijet invariant mass provides discrimination between signal and background and 
is a critical variable used in the multivariate analysis as described below.
Improvement of the dijet mass resolution directly results in an improvement of the $WH$ signal sensitivity.
To improve dijet invariant mass resolution, we developed a neural network 
$b$-jet energy correction method. The neural network was trained using a sample of Monte Carlo simulated 
$WH\rightarrow l\nu b\bar b$ events. During training, jet observables were used as input values, and
the energy of the corresponding $b$ quark was used as the target value. 

For each jet, we studied 40 variables related to the calorimeter energy, the charged tracks, and 
the displaced vertices within the jet cone of 0.4 and converged on nine well-modeled input variables most 
optimal for the jet-energy correction. The four calorimeter variables chosen are the jet $E_T$ before and
after the standard jet correction, jet $p_T$, and jet transverse mass. The tracking variables chosen are 
the sum $p_T$ and the maximum $p_T$ of the set of tracks within the jet cone. For the jet tagged 
by SECVTX we also include the vertex variables such as the secondary 
vertex transverse decay length, its uncertainty, and fitted secondary vertex $p_T$.  
Further details can be found in Ref.~\cite{Timo_NIM}. Without (with) applying NN corrections to $b$-jets in the 
Higgs decays, the dijet mass resolution is $\sim$15\% ($\sim$11\%) for double-tagged events, 
and $\sim$17\% ($\sim$13\%) for single-tagged events.

\subsection{Bayesian Neural Network Discriminant}
\label{sec:bnnd}
To improve the signal-to-background discrimination further, we employed a Bayesian neural network 
(BNN) trained on a variety of kinematic variables to distinguish $WH$ events from the 
background~\cite{BNN,BNNbook}. 
%The application of the Bayesian learning approach to neural networks results 
%in a flexible and powerful neural network that is more stable and less prone to over-training. 
%Within BNN, all sources of uncertainty are expressed and measured by probabilities. 
For this analysis, we employ distinct BNN discriminant functions that were optimized separately for the 
different tagging categories and each Higgs boson mass in order to maximize the sensitivity. 

The BNN configuration has N input variables, 2N hidden nodes, and one output node. The input variables were 
selected by an iterative BNN optimization procedure from a large number of possible variables. The 
optimization procedure identified the most sensitive one-variable BNN, then looped over all remaining variables
and found the most sensitive two-variable BNN. The process continued until adding a new variable no longer improved
the sensitivity. The discriminant then is used to do hypothesis testing 
of a $WH$ signal in the simulated data as a function of Higgs mass, which improves the background rejection with a sensitivity gain of
25\% compared to the most sensitive variable alone.

The discriminant used for the ST+ST tag category is trained using N=7 input variables. The most sensitive variable
 is $M_{jj}$, the invariant mass calculated from the two tight jets after using the neural-network-based 
jet energy correction as described in Sec.~\ref{sec:bjec}. The second input variable is the $p_T$ 
imbalance, which is the difference between the scalar sum of the $p_T$ of all measured objects and the {\MET},
$p_T$(jet1) + $p_T$(jet2) + $p_T$(lep) $-$~{\MET}. 
The third variable, $M^{max}_{l\nu j}$, is the invariant mass of the lepton, {\MET}, and one of the two jets,
where the jet is chosen to give the maximum invariant mass. The fourth variable is $Q_{lep} \times \eta_{lep}$, 
the signed product of the electric charge times the $\eta$ of the charged lepton. The fifth variable is $\Sigma E_T$(loose jets), 
which is the scalar sum of loose jets transverse energy. A loose jet is defined as a jet having $|\eta|<2.4$,
$E_T>12$ GeV, but failing the tight-jet requirement ($E_T>20$ GeV and $|\eta|<2.0$). 
The sixth variable is the $p_T$ of the reconstructed $W$.
The last variable is $H_T$, the scalar sum of the event transverse energies $H_T=\Sigma E_T$ (jets) +
$p_T$(lepton) +~{\MET}.   

The discriminant used for both the ST+JP and ST+NN tag categories is trained with the same input variables as the 
ST+ST category, except that the variable $M^{max}_{l\nu j}$ is replaced with $M^{min}_{l\nu j}$ and 
the $p_T$ imbalance is replaced with the {\MET}. The discriminant used for the single ST tag category is trained
with
the same input variables as the ST+ST category with the exception that $M^{max}_{l\nu j}$ is replaced by {\MET} and 
an extra variable is added. The new variable is the output of an artificial-neural-network-based heavy flavor 
separator trained to distinguish $b$-quark jets from the charm and light flavor jets 
after SECVTX tagging~\cite{singletop}. 
Distributions of all these variables are checked for both the pretag and tagged sample to ensure that they are 
described well by the background model. 
%The simulated background events match the real data well. 

\begin{figure*}[htbp]
   \begin{center}
     \includegraphics[width=8.0cm]{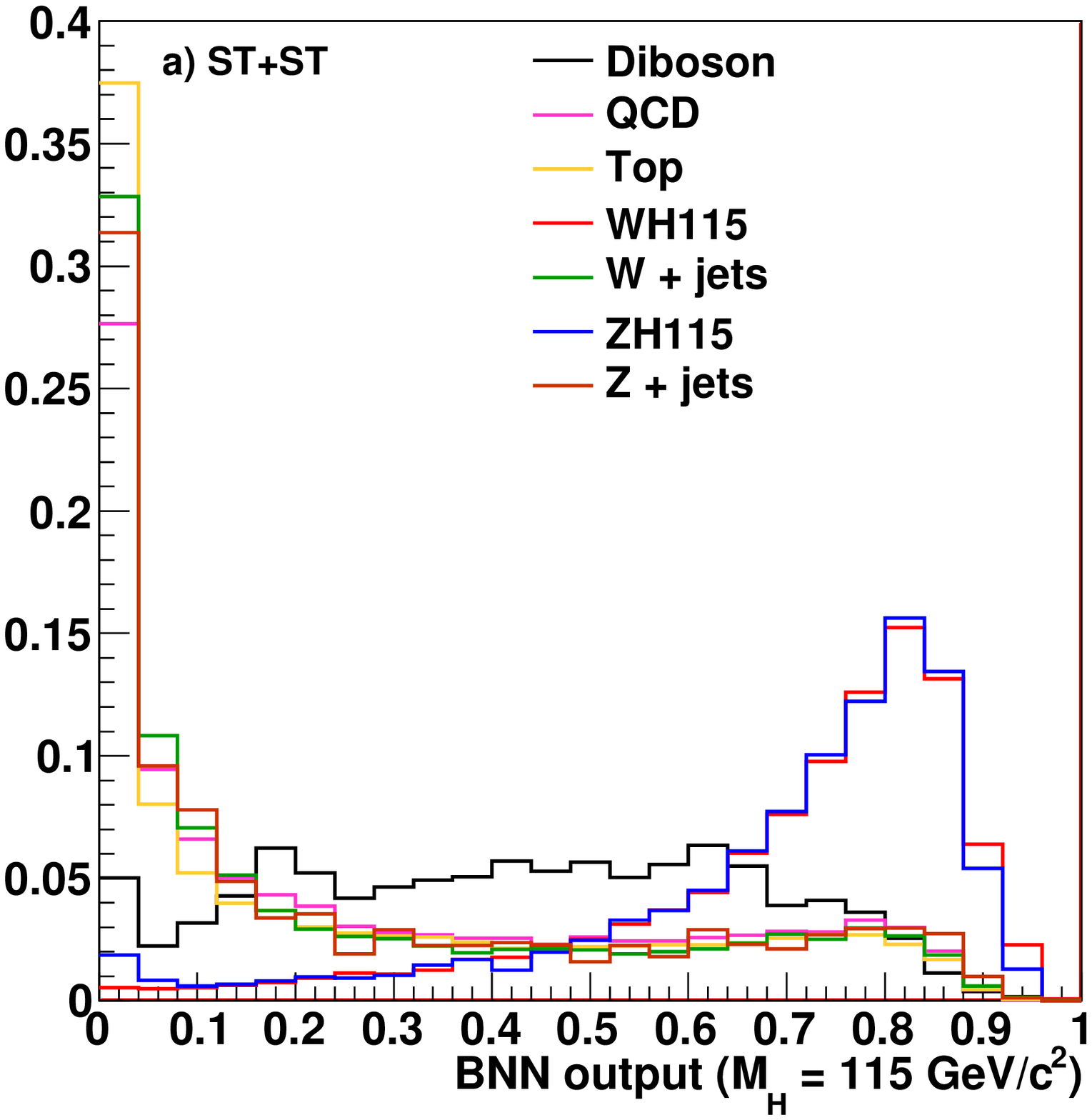}
     \includegraphics[width=8.0cm]{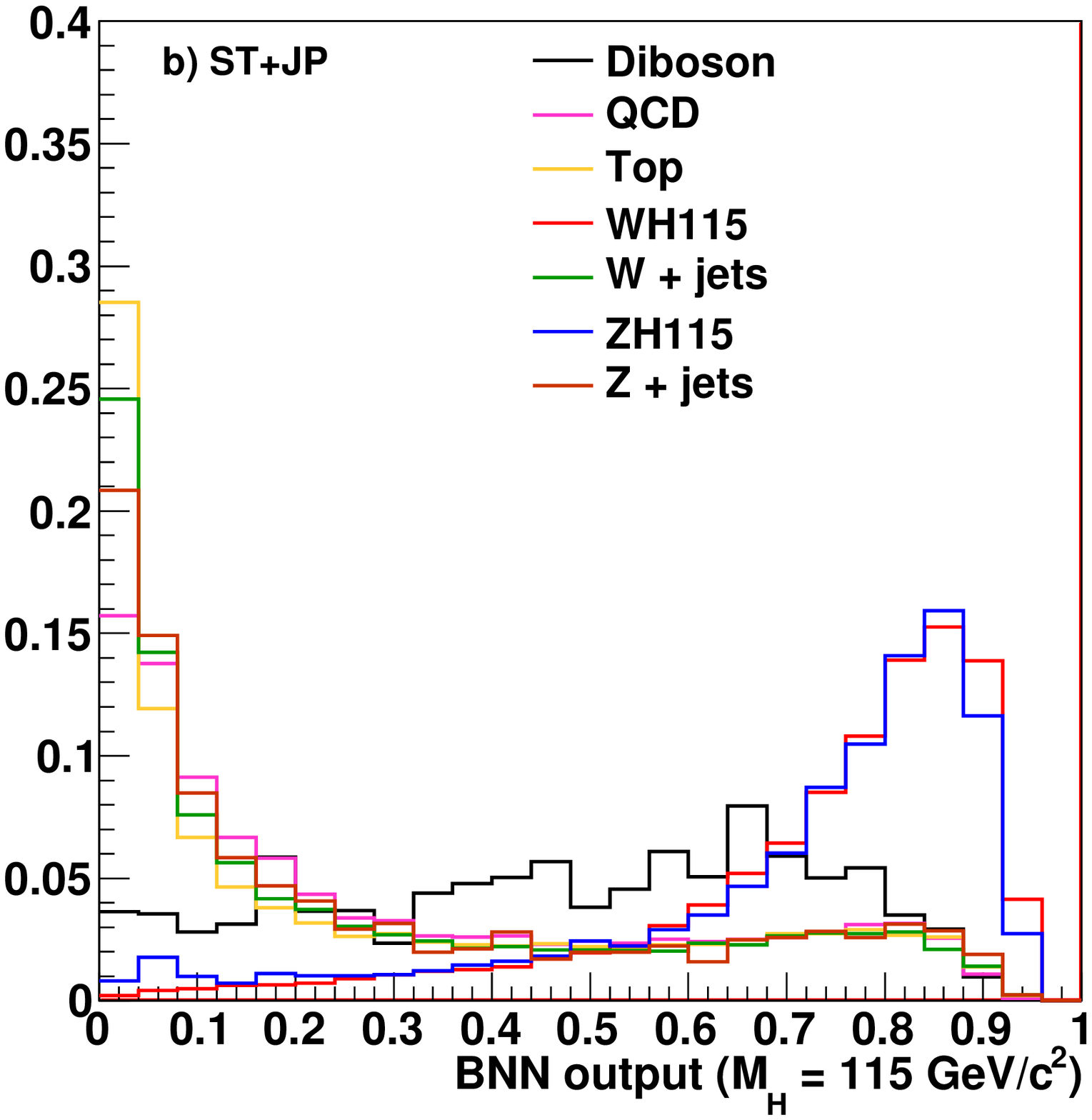}
     \includegraphics[width=8.0cm]{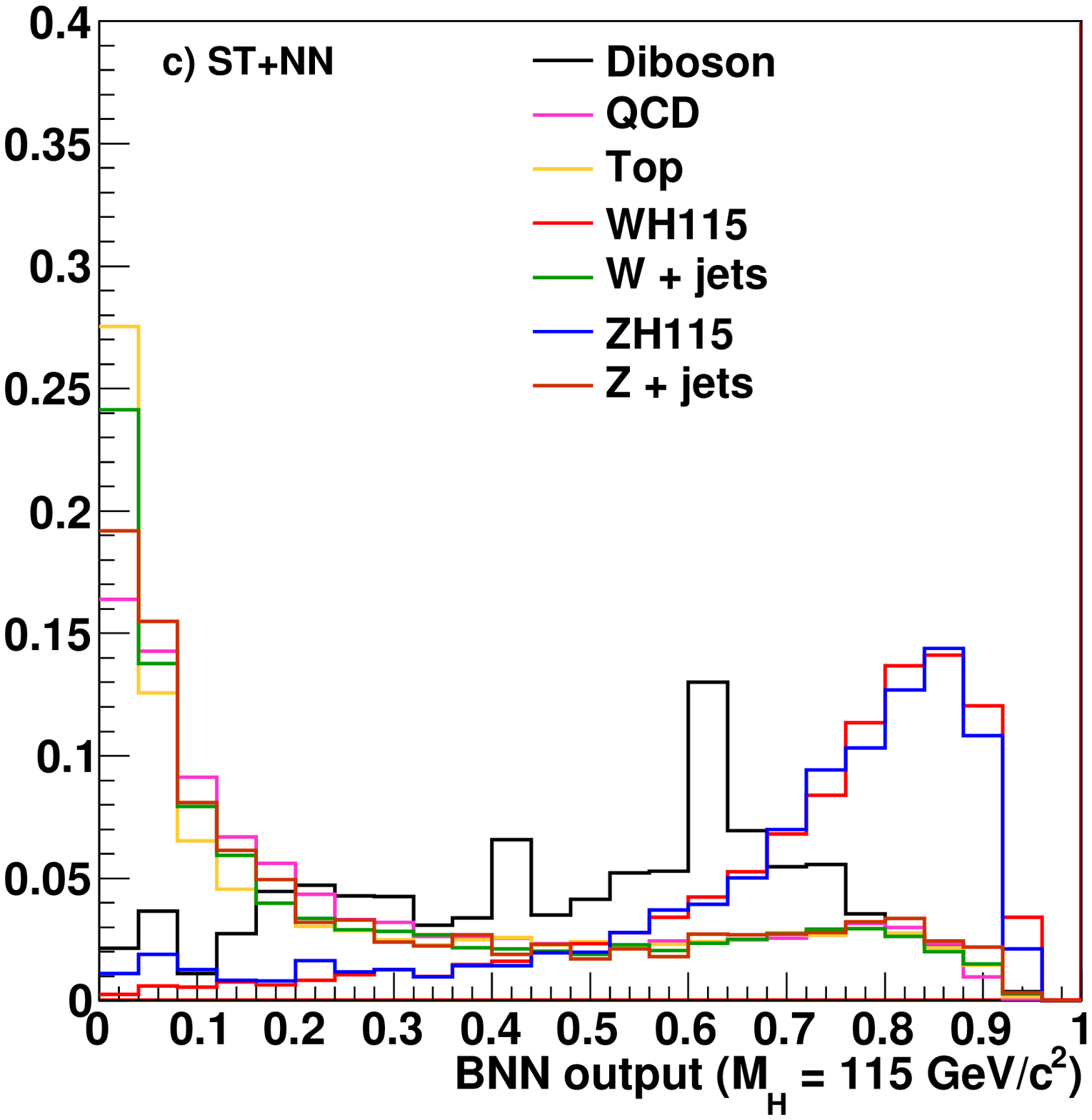}
     \includegraphics[width=8.0cm]{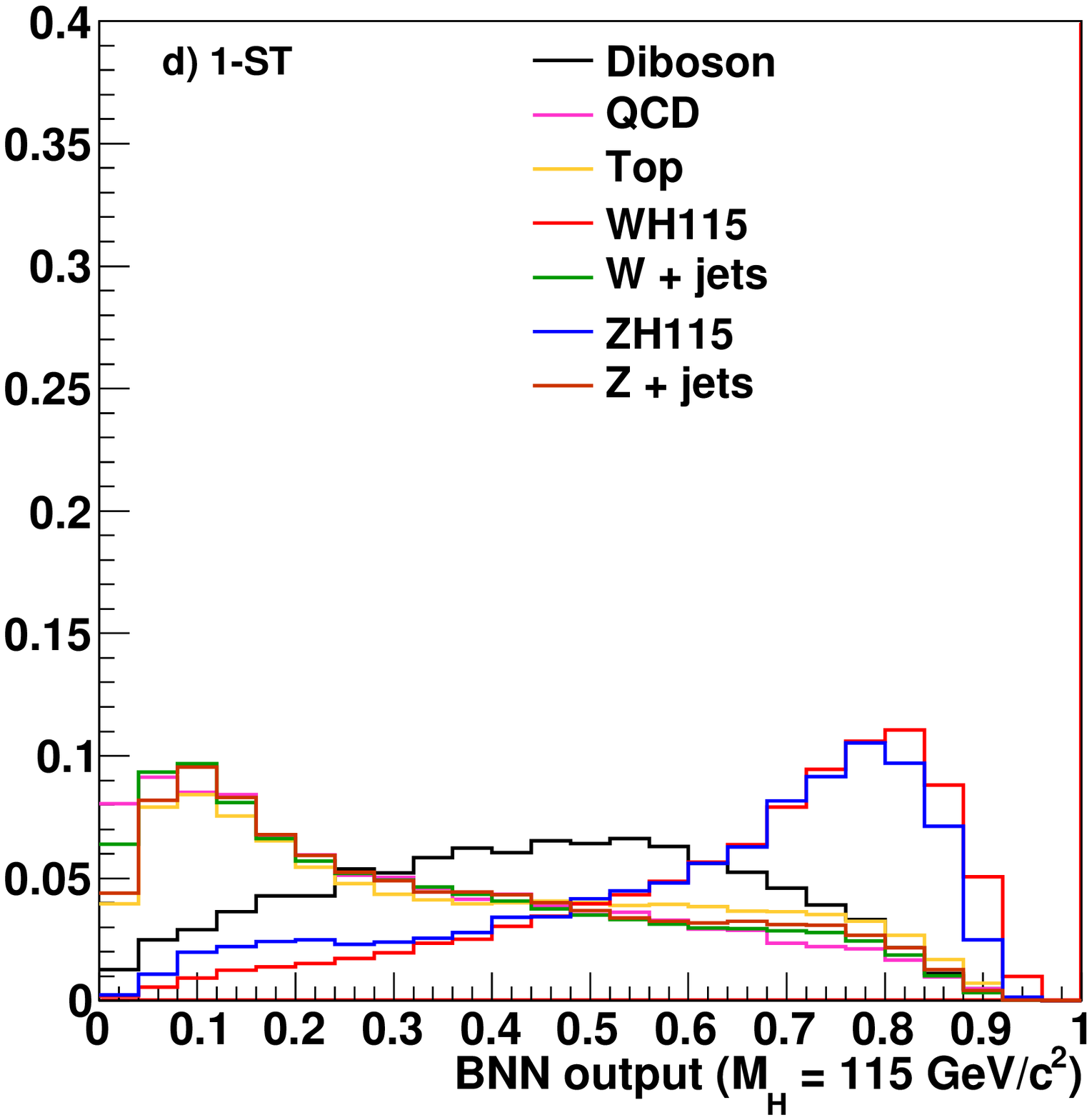}
     \caption{Comparison of the BNN output for signal ($M_H=115$ GeV/c$^2$) and background events 
with all lepton types included. From (a) - (d) the $b$-tag categories are ST+ST, ST+JP, ST+NN, and ST, respectively. 
Signal and background histograms are each normalized to unit area. The $WH$ and $ZH$ signals peak near the 1.0 value.
The QCD multijet, top quark, $W$ + jets and $Z$ + jets peak near the 0.0 value. The diboson background has a broad peak 
in the middle region as its kinematics is very close to the signal ones. The diboson spike in figure (c) is a 
statistical fluctuation.}
     \label{fig:NNnormout}
   \end{center}
 \end{figure*}

The training is defined such that the neural network attempts to produce an output as close to 1.0 as possible
for the Higgs boson signal events and as close to 0.0 as possible for background events.  
Figure~\ref{fig:NNnormout} shows a shape comparison of the BNN output between signal and background events for 
the ST+ST, ST+JP, ST+NN, and ST sample, respectively. 

%%%%%%%%%%%%%%%%%%%%%
% Section 8 Results %
%%%%%%%%%%%%%%%%%%%%%
%\input{result}
\section{Results}\label{sec:results}
\begin{figure*}[htbp]
   \begin{center}
     \includegraphics[width=8.0cm]{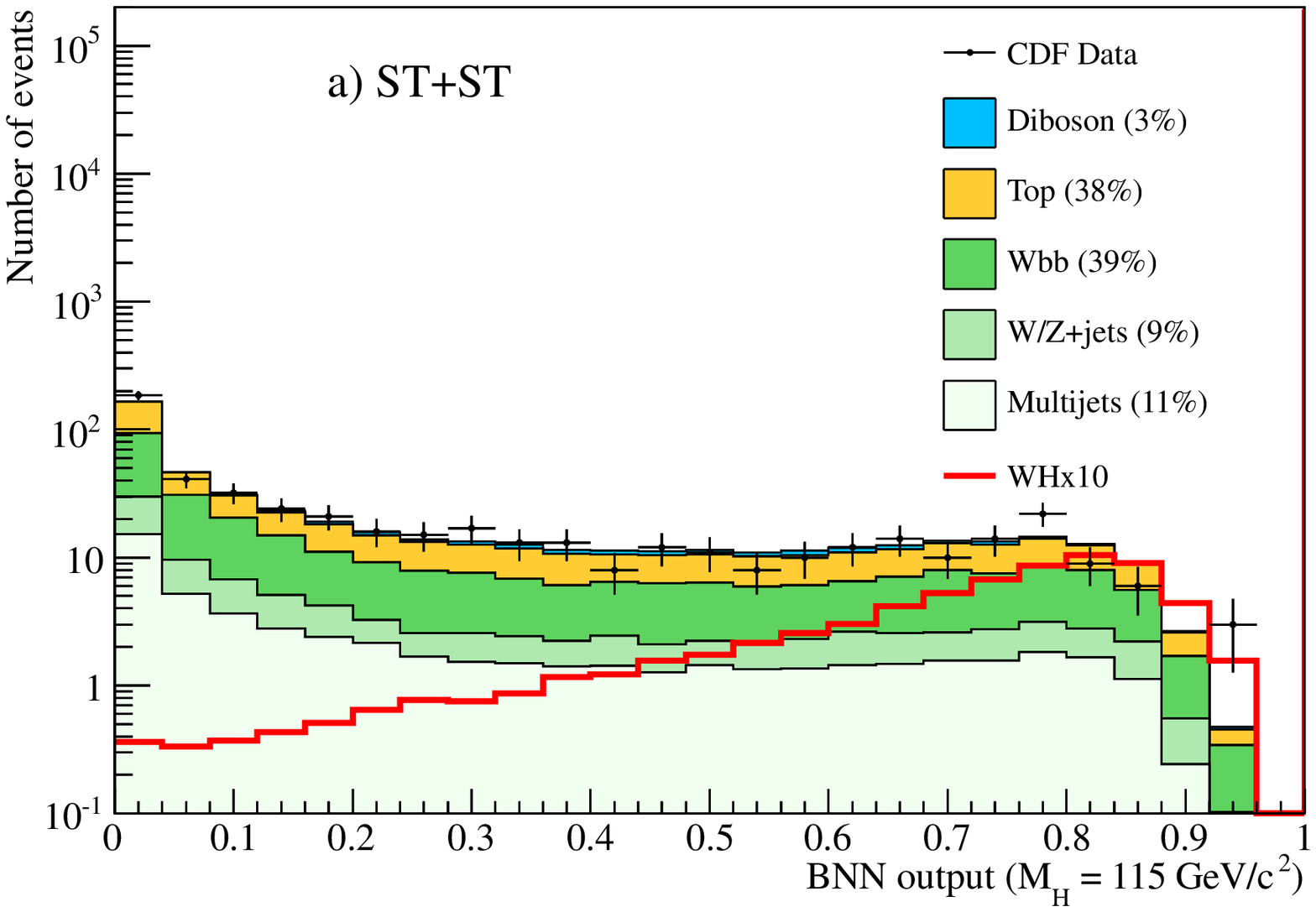}
     \includegraphics[width=8.0cm]{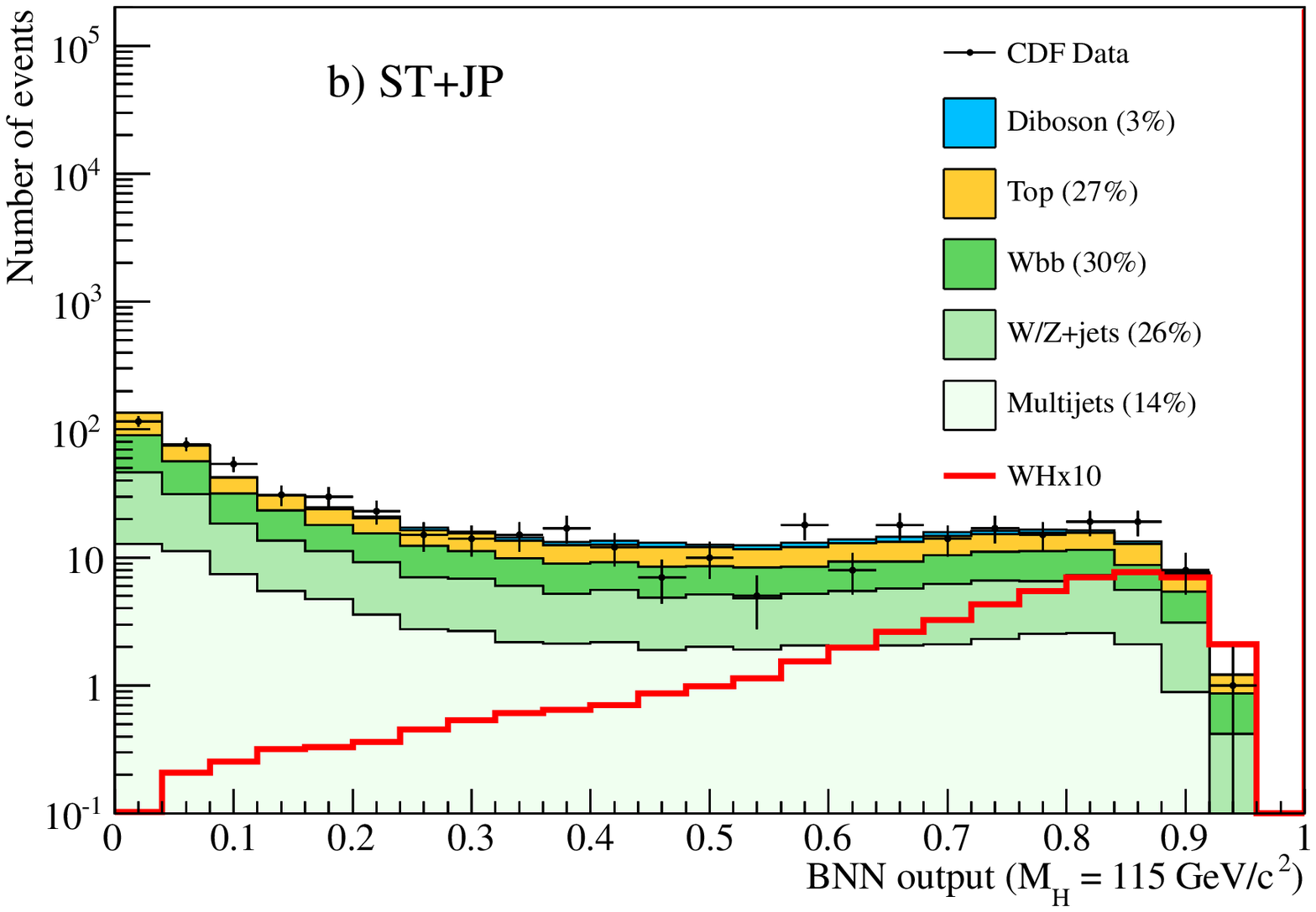}
     \includegraphics[width=8.0cm]{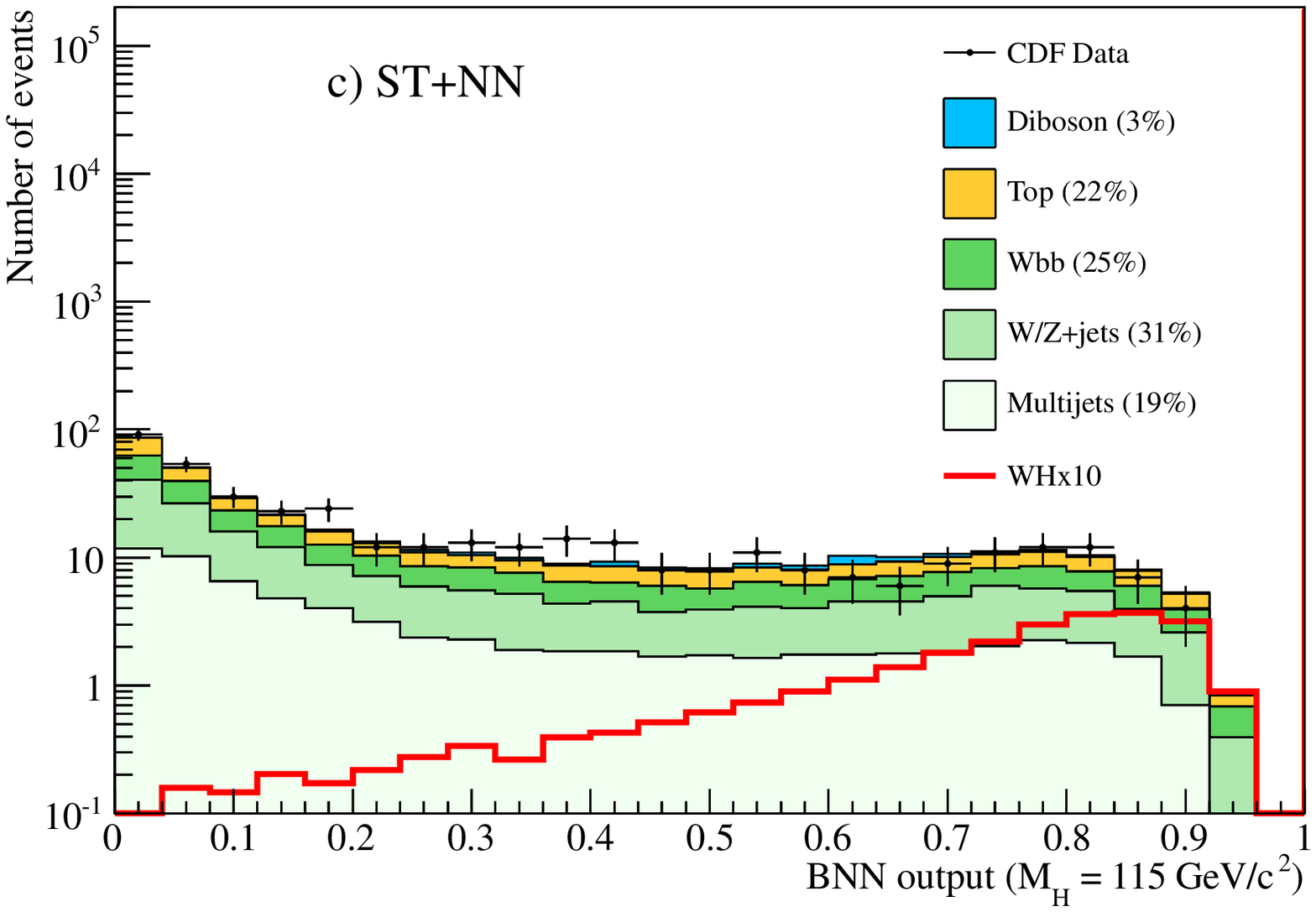}
     \includegraphics[width=8.0cm]{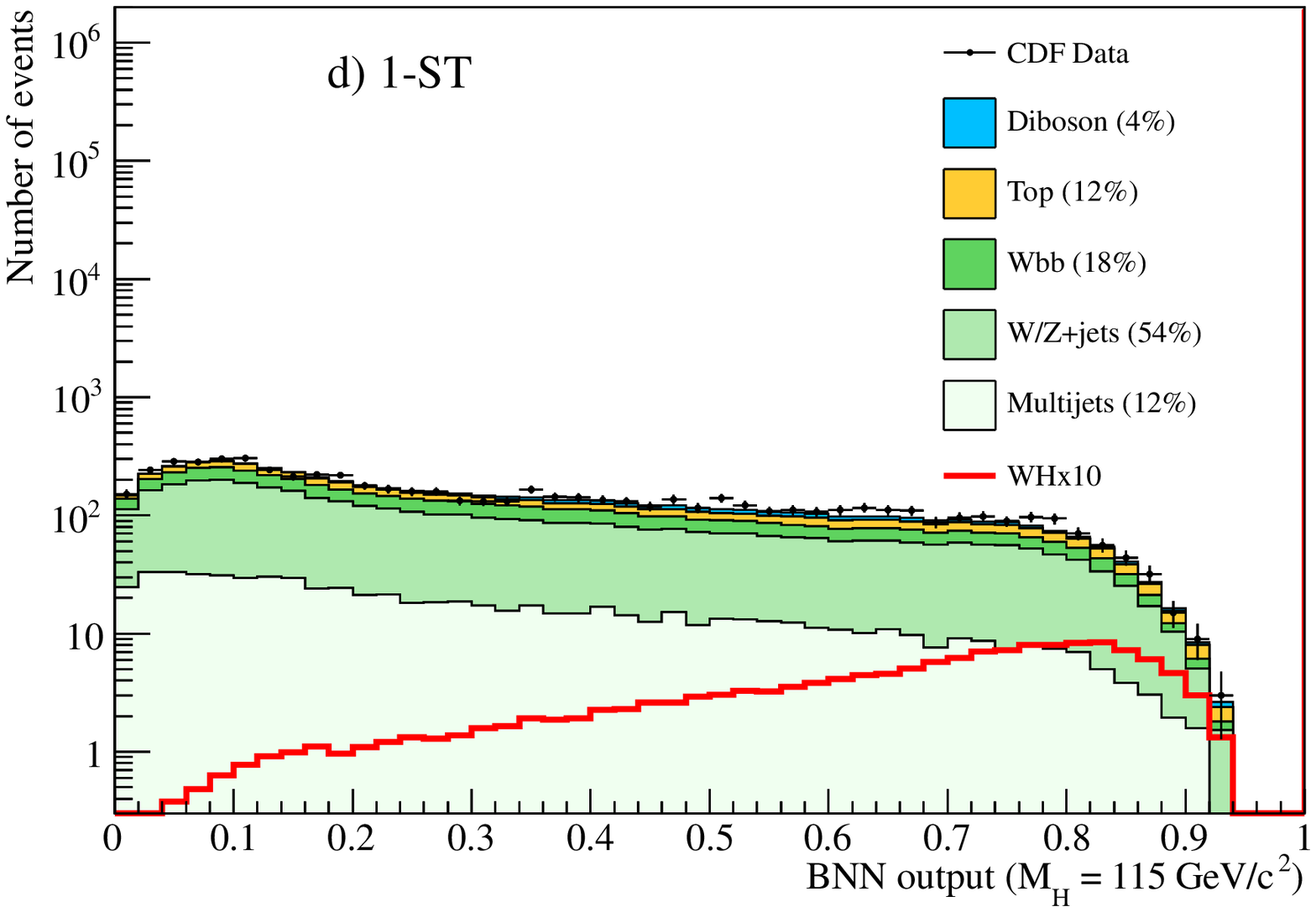}
     \caption{The observed data and predicted BNN output for signal ($M_H=115$ GeV/c$^2$) and background events with all 
leptons included. From (a) - (d) the $b$-tag categories are ST+ST, ST+JP, ST+NN, and ST, respectively.} 
     \label{fig:NNoutput}
   \end{center}
 \end{figure*}

We perform a direct search for an excess in the signal region of the BNN output distribution from double-tagged and single-tagged $W$ + 2 jets 
events. Figure~\ref{fig:NNoutput} shows the BNN output distributions for each $b$-tagging category. The data and background predictions are 
in good agreement. 

We use a binned likelihood fit~\cite{jason,trj} to the observed BNN output 
distributions to test the presence of a $WH$ signal. For optimal sensitivity, we perform a 
simultaneous search in each $b$-tag and lepton category. The total likelihood is the product
of the single Poisson likelihoods used in each independent sample. 
The likelihood fit accommodates the uncertainties on our background estimate by letting the overall background 
prediction float within Gaussian constraints. The systematic uncertainties associated with the shape of the 
BNN output due to JES uncertainty are also included for both signal and background. We use a different set of 
background and signal BNN template shapes for each combination of lepton type and tag category. 
We correlate the systematic uncertainties appropriately across different lepton types and tag 
categories. We find no evidence for a Higgs boson signal in our sample. We use Bayesian limits with a positive
flat prior and set 95\% C.L. upper limits on the $WH$ cross section times
branching ratio, $\sigma(p\bar p \rightarrow W^{\pm}H)\cdot {\cal B}(H\rightarrow b\bar b)$, relative to 
the rate predicted by the standard model. 

\begin{figure*}[htbp]
  \begin{center}
    \includegraphics[width=.45\textwidth]{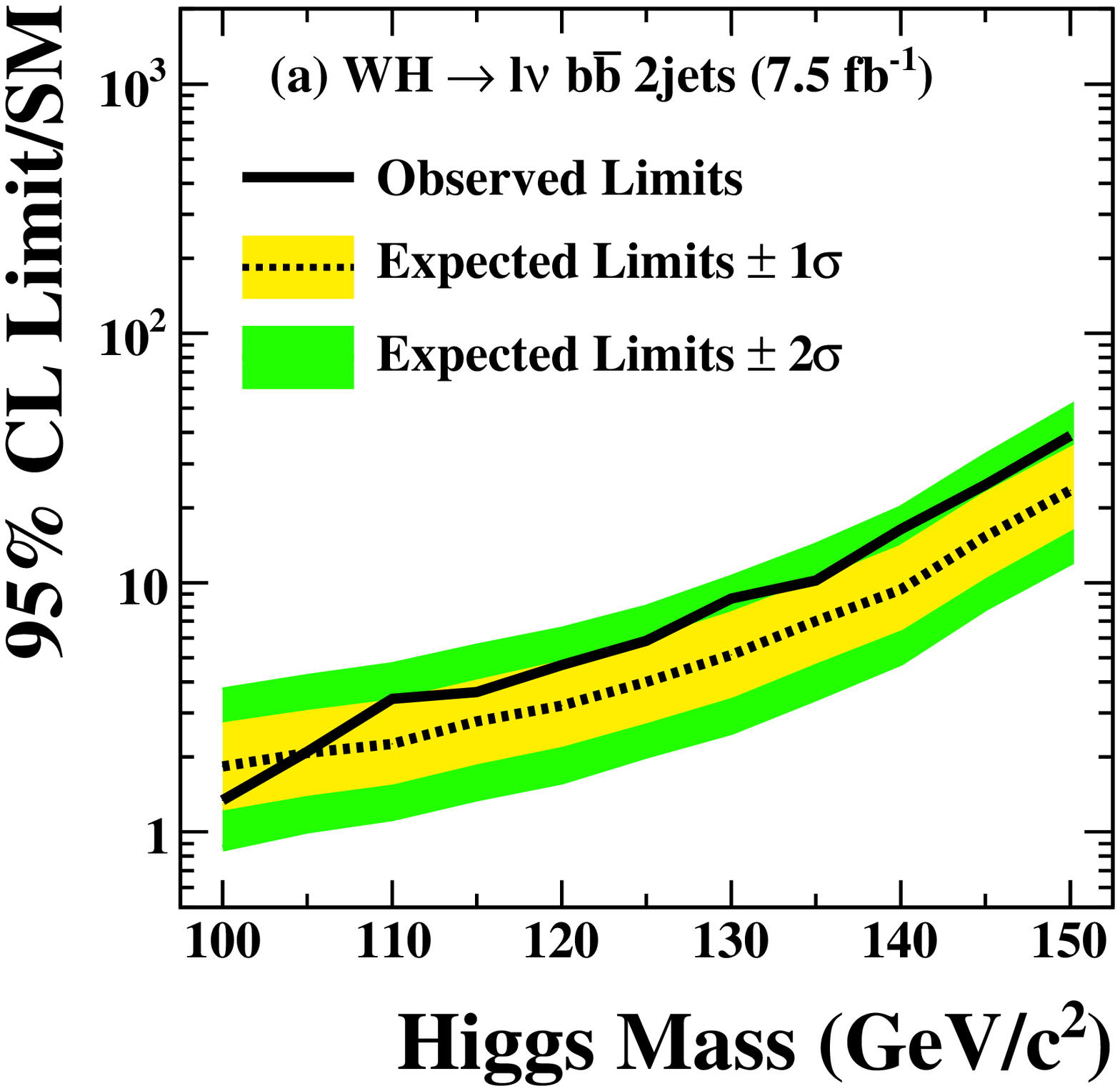}
    \includegraphics[width=.45\textwidth]{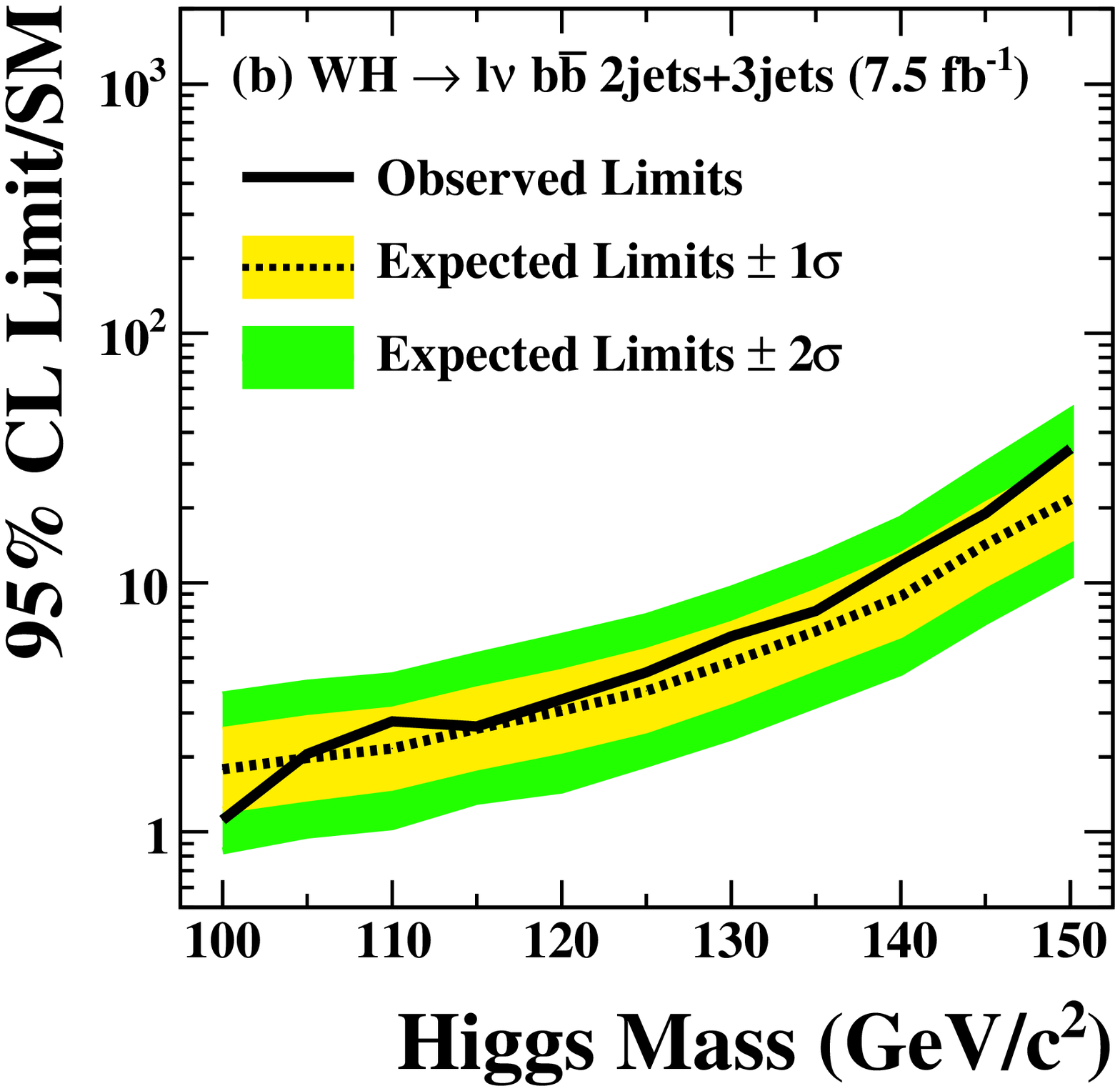}
    \caption{Observed and expected upper limits at 95\% C.L. on Higgs boson production times branching ratio with 
respect to the SM expectation for all lepton and tag categories combined as a function of the Higgs boson mass 
for the present analysis (a) and after 
   combination with the independent three jet analysis with a matrix element (b).}
    \label{fig:Limit}
  \end{center}
\end{figure*}

We compare our observed limits to our expected sensitivity by generating statistical trials according to the 
background-only model and analyzing them as our data.
The combined expected and observed limits for 
all the lepton types are shown in Figure~\ref{fig:Limit} and Table~\ref{table:Limit}. 
Limits are also determined for the combination of this analysis with  
the independent $WH$ search using a matrix element technique for events with three jets~\cite{me}.  
The luminosity used in the three jet analysis is 5.6 fb$^{-1}$. The combination improves  
the expected $WH$ sensitivity by about 5\%  over the $W$ + two jet result alone. 
The observed limits in the two jet channel in the mass range above $m_H>110$ GeV/c$^2$ are one standard deviation
 higher than expected.
After combining with three jet bin, our limits become closer to the expectation. 

\begin{table*}
\begin{center}
  \caption{Observed and expected upper limits at 95\% C.L. normalized to the SM expectation on 
    $\sigma (p\bar{p} \rightarrow WH) \times {\cal B} (H \to b\bar{b})$ as a function of Higgs mass, 
    including all lepton and tag categories, in the presented analysis and after combination with an 
    independent search using a matrix element analysis for events with three jets. }
  \begin{tabular}{ccccc}
    % skip a line with no table stuff
    \hline
    \hline
    \multicolumn{5}{c}{Upper Limits/SM for Combined Lepton and Tag Categories} \\ \hline 
   & \multicolumn{2}{c}{ $W$ + 2 jets } & \multicolumn{2}{c}{ $W$ + 2, 3 jets } \\ \hline
    $m_H$ ($\gevcc$)  & Observed & Expected & Observed & Expected \\
    \hline
    100 & 1.34  & 1.83  & 1.12 & 1.79  \\
    105 & 2.10  & 2.08  & 2.06 & 1.98 \\
    110 & 3.42  & 2.26  & 2.78 & 2.17 \\
    115 & 3.64  & 2.78  & 2.65 & 2.60 \\
    120 & 4.68  & 3.22  & 3.40 & 3.06 \\
    125 & 5.84  & 4.01  & 4.36 & 3.69 \\
    130 & 8.65  & 5.13  & 6.09 & 4.80 \\
    135 & 10.2  & 7.02  & 7.71 & 6.40 \\
    140 & 16.4  & 9.39  & 12.3 & 8.84 \\
    145 & 24.7  & 15.3  & 18.9 & 14.2 \\
    150 & 38.8  & 23.4  & 34.4 & 21.6 \\
    \hline
  \end{tabular}
  \label{table:Limit}
\end{center}
\end{table*}

This $WH(ZH) \to \ell \nu (\ell \ell \!\!\!\!\!\!\not\,\,\,\,) b\bar{b}$ analysis represents a substantial improvement in sensitivity over the prior
analysis using a neural network~\cite{jason}.
The increase in sensitivity is 25\% at $m_{H}$ = 115 GeV/c${^2}$ in addition to the improvement from a larger sample size, and is 
mainly due to 
the improvement of analysis techniques that include the BNN discriminant, the $b$-jet energy correction, 
the additional {\MET} triggers, the loose leptons, and the optimized $b$-tagging strategies.

%%%%%%%%%%%%%%%%%%%%%%%%
% Section 9 Conclusion %
%%%%%%%%%%%%%%%%%%%%%%%%
%\input{conclusion}
\section{Conclusions}\label{sec:conclusions}
We have presented the results of a CDF search for the standard model Higgs boson decaying to $b\bar b$ final states,
produced in association
with a $W$ boson decaying into a charged lepton and neutrino. We find that for the dataset corresponding to an integrated 
luminosity of 7.5 fb$^{-1}$, the data agree with the SM background predictions.
We therefore set upper limits on the Higgs boson production cross section times the $H\rightarrow b\bar b$ branching ratio
with respect to the standard model prediction. 
For the mass range of 100 GeV/c$^2$ through 150 GeV/c$^2$ we set observed (expected) upper limits at 95\% C.L. from 
1.34 (1.83) to 38.8 (23.4). For 115 GeV/c$^2$ the upper limit is 3.64 (2.78). When we combine 
this search with an independent search using events with three jets~\cite{me}, we set more stringent limits in the same mass range 
from 1.12 (1.79) to 34.4 (21.6).  For 115 and 125 GeV/c$^2$ the upper limits are 2.65 (2.60) and 4.36 (3.69), 
respectively. Improved analysis techniques
have resulted in an increase in sensitivity over the previous 2.7 fb$^{-1}$ analysis~\cite{jason} by
25\% more than the expectation from simple luminosity scaling.

The search results in this channel at the CDF experiment are the most sensitive low-mass Higgs boson search at the Tevatron. 
While the LHC experiments will continue to improve their sensitivity to the low-mass Higgs boson, which is obtained 
primarily from searches 
in the diphoton final state, we expect that the searches in the $H\rightarrow b\bar b $ channel at the Tevatron will provide a crucial
test on the existence and nature of the low-mass Higgs boson.

%%%%%%%%%%%%%%%%%%%%%%%%%%%%%%%%%%%%%%%%%%%%%%%%%%
% Acknowledgements
%%%%%%%%%%%%%%%%%%%%%%%%%%%%%%%%%%%%%%%%%%%%%%%%%%
\begin{acknowledgments}
We thank the Fermilab staff and the technical staffs of the participating institutions for their vital contributions. 
This work was supported by the U.S. Department of Energy and National Science Foundation; the Italian Istituto Nazionale 
di Fisica Nucleare; the Ministry of Education, Culture, Sports, Science and Technology of Japan; the Natural Sciences and 
Engineering Research Council of Canada; the National Science Council of the Republic of China; 
the Swiss National Science Foundation; the A.P. Sloan Foundation; the Bundesministerium f\"ur Bildung und Forschung, Germany; 
the Korean World Class University Program, the National Research Foundation of Korea; 
the Science and Technology Facilities Council and the Royal Society, UK; the Russian Foundation for Basic Research; 
the Ministerio de Ciencia e Innovaci\'{o}n, and Programa Consolider-Ingenio 2010, Spain; the Slovak R\&D Agency; 
the Academy of Finland; and the Australian Research Council (ARC).
\end{acknowledgments}

%\bibliography{reference}

\begin{thebibliography}{36}
\expandafter\ifx\csname natexlab\endcsname\relax\def\natexlab#1{#1}\fi
\expandafter\ifx\csname bibnamefont\endcsname\relax
  \def\bibnamefont#1{#1}\fi
\expandafter\ifx\csname bibfnamefont\endcsname\relax
  \def\bibfnamefont#1{#1}\fi
\expandafter\ifx\csname citenamefont\endcsname\relax
  \def\citenamefont#1{#1}\fi
\expandafter\ifx\csname url\endcsname\relax
  \def\url#1{\texttt{#1}}\fi
\expandafter\ifx\csname urlprefix\endcsname\relax\def\urlprefix{URL }\fi
\providecommand{\bibinfo}[2]{#2}
\providecommand{\eprint}[2][]{\url{#2}}

\bibitem[{\citenamefont{Higgs}(1964{\natexlab{a}})}]{Higgs:1964ia}
\bibinfo{author}{\bibfnamefont{P.~W.} \bibnamefont{Higgs}},
  \bibinfo{journal}{Phys. Lett.} \textbf{\bibinfo{volume}{12}},
  \bibinfo{pages}{132} (\bibinfo{year}{1964}{\natexlab{a}}).

\bibitem[{\citenamefont{Higgs}(1964{\natexlab{b}})}]{Higgs:1964pj}
\bibinfo{author}{\bibfnamefont{P.~W.} \bibnamefont{Higgs}},
  \bibinfo{journal}{Phys. Rev. Lett.} \textbf{\bibinfo{volume}{13}},
  \bibinfo{pages}{508} (\bibinfo{year}{1964}{\natexlab{b}}).

\bibitem[{\citenamefont{Guralnik}(1964{\natexlab{b}})}]{Higgs:1964ghk}
\bibinfo{author}{\bibfnamefont{G.~S.} \bibnamefont{Guralnik}},
\bibinfo{author}{\bibfnamefont{C.~R.} \bibnamefont{Hagen}},
\bibnamefont{and} \bibinfo{author}{\bibfnamefont{T.~W.~B.} \bibnamefont{Kibble}},
  \bibinfo{journal}{Phys. Rev. Lett.} \textbf{\bibinfo{volume}{13}},
  \bibinfo{pages}{585} (\bibinfo{year}{1964}{\natexlab{b}}).

\bibitem[{\citenamefont{Englert}(1964{\natexlab{b}})}]{Higgs:1964eb}
\bibinfo{author}{\bibfnamefont{F.} \bibnamefont{Englert}}
\bibnamefont{and} \bibinfo{author}{\bibfnamefont{R.} \bibnamefont{Brout}},
  \bibinfo{journal}{Phys. Rev. Lett.} \textbf{\bibinfo{volume}{13}},
  \bibinfo{pages}{321} (\bibinfo{year}{1964}{\natexlab{b}}).

\bibitem[{\citenamefont{Barate et~al.}(2003)}]{Barate:2003sz}
\bibinfo{author}{\bibfnamefont{R.}~\bibnamefont{Barate}} \bibnamefont{{\it et~al.}}
  (\bibinfo{collaboration}{LEP Working Group for Higgs boson searches}),
  \bibinfo{journal}{Phys. Lett. B} \textbf{\bibinfo{volume}{565}},
  \bibinfo{pages}{61} (\bibinfo{year}{2003}).

\bibitem[{\citenamefont{In this paper, lepton ($\ell$) denotes electron ($e^\pm$), 
muon ($\mu^\pm$), or small contribution from one-prong tau ($\tau^\pm$) decay, and $\nu$ denotes each 
corresponding neutrinos}}]{leptonNote}
\bibinfo{author}{\bibnamefont{In this paper, lepton ($\ell$) denotes electron ($e^\pm$), 
muon ($\mu^\pm$), or small contribution from one-prong tau ($\tau^\pm$) decay, and $\nu$ denotes each 
corresponding neutrinos}}.

\bibitem[{The Tevatron New Phenomena and Higgs Working Group (2011)}]{TeVcombination11}
\bibinfo{author}{\bibnamefont{{The Tevatron New Phenomena and Higgs Working Group (2011)}}}, \eprint{arXiv:1107.5518}.

\bibitem[{\citenamefont{Aad et~al.}(2012)}]{atlas}
\bibinfo{author}{\bibfnamefont{G.} \bibnamefont{Aad}}
\bibnamefont{{\it et~al.}} (\bibinfo{collaboration}{ATLAS Collaboration}), 
  \bibinfo{journal}{Phys. Lett. B
  } \textbf{\bibinfo{volume}{710}}, \bibinfo{pages}{49}
  (\bibinfo{year}{2012}).

\bibitem[{\citenamefont{Chatrchyan et~al.}(2012)}]{cms}
\bibinfo{author}{\bibfnamefont{S.} \bibnamefont{Chatrchyan}}
\bibnamefont{{\it et~al.}} (\bibinfo{collaboration}{CMS Collaboration}), 
\bibinfo{journal}{Phys. Lett. B} \textbf{\bibinfo{volume}{710}}, 
\bibinfo{pages}{26} (\bibinfo{year}{2012}).

\bibitem[{\citenamefont{http://lepewwg.web.cern.ch/LEPEWWG/}()}]{EWGfits2009Summer}
\bibinfo{author}{\bibnamefont{The LEP Electroweak Working Group}},
\bibinfo{howpublished}{http://lepewwg.web.cern.ch/LEPEWWG/},\eprint(arXiv:0911.2604).

\bibitem[{\citenamefont{{Tev4LHC Higgs Working Group}}()}]{Tev4LHC}
\bibinfo{author}{\bibnamefont{{Tev4LHC Higgs Working Group}}},
  \emph{\bibinfo{title}{{Standard Model Higgs cross sections at hadron
  colliders}}},
  \bibinfo{howpublished}{http://maltoni.home.cern.ch/maltoni/TeV4LHC/SM.html}.

\bibitem[{\citenamefont{Djouadi et~al.}(1998)\citenamefont{Djouadi, Kalinowski,
  and Spira}}]{Djouadi:1997yw}
\bibinfo{author}{\bibfnamefont{A.}~\bibnamefont{Djouadi}},
  \bibinfo{author}{\bibfnamefont{J.}~\bibnamefont{Kalinowski}},
  \bibnamefont{and} \bibinfo{author}{\bibfnamefont{M.}~\bibnamefont{Spira}},
  \bibinfo{journal}{Comput. Phys. Commun.} \textbf{\bibinfo{volume}{108}},
  \bibinfo{pages}{56} (\bibinfo{year}{1998}).


\bibitem[{\citenamefont{Aaltonen et~al.}(2011)}]{jason}
\bibinfo{author}{\bibfnamefont{T.}~\bibnamefont{Aaltonen}}
  \bibnamefont{{\it et~al.}} (\bibinfo{collaboration}{CDF Collaboration}), 
  \bibinfo{journal}{Phys. Rev. D
  } \textbf{\bibinfo{volume}{85}}, \bibinfo{pages}{052002}
  (\bibinfo{year}{2012}).

\bibitem[{\citenamefont{Aaltonen et~al.}(2011)}]{me}
\bibinfo{author}{\bibfnamefont{T.}~\bibnamefont{Aaltonen}}
  \bibnamefont{{\it et~al.}} (\bibinfo{collaboration}{CDF Collaboration}), 
  \bibinfo{journal}{Phys. Rev. D 
  } \textbf{\bibinfo{volume}{85}}, \bibinfo{pages}{072001}
  (\bibinfo{year}{2012}).

\bibitem[{\citenamefont{Abazov et~al.}(2011)}]{d0}
\bibinfo{author}{\bibfnamefont{V.~M.}~\bibnamefont{Abazov}}
  \bibnamefont{{\it et~al.}} (\bibinfo{collaboration}{D0 Collaboration}), 
  \bibinfo{journal}{Phys. Lett. B 
  } \textbf{\bibinfo{volume}{698}}, \bibinfo{pages}{6}
  (\bibinfo{year}{2011}).

\bibitem[{\citenamefont{Neal}()}]{BNN}
\bibinfo{author}{\bibfnamefont{R.~M.} \bibnamefont{Neal}},
  \bibinfo{journal}{http://www.cs.toronto.edu/\~radford/fbm.software.html}.

\bibitem[{\citenamefont{Neal}(1996)}]{BNNbook}
\bibinfo{author}{\bibfnamefont{R.~M.} \bibnamefont{Neal}},
  \emph{\bibinfo{title}{{Bayesian Learning for Neural Networks}}}
  (\bibinfo{publisher}{Springer}, \bibinfo{year}{1996}).

\bibitem[{\citenamefont{Acosta et~al.}(2005{\natexlab{b}})}]{Acosta:2004yw}
\bibinfo{author}{\bibfnamefont{D.}~\bibnamefont{Acosta}} \bibnamefont{{\it et~al.}}
  (\bibinfo{collaboration}{CDF Collaboration}), \bibinfo{journal}{Phys. Rev. D}
  \textbf{\bibinfo{volume}{71}}, \bibinfo{pages}{032001}
  (\bibinfo{year}{2005}{\natexlab{b}}).

\bibitem[{\citenamefont{Sill}(2000)}]{Sill:2000svx}
\bibinfo{author}{\bibfnamefont{A.}~\bibnamefont{Sill}} 
(\bibinfo{collaboration}{CDF Collaboration}),
  \bibinfo{journal}{Nucl. Instrum. Methods A} \textbf{\bibinfo{volume}{447}},
  \bibinfo{pages}{1} (\bibinfo{year}{2000}).

\bibitem[{\citenamefont{Affolder}(2004)}]{Affolder:2004cot}
\bibinfo{author}{\bibfnamefont{T.}~\bibnamefont{Affolder}} \bibnamefont{{\it et~al.}},
  \bibinfo{journal}{Nucl. Instrum. Methods A} \textbf{\bibinfo{volume}{526}},
  \bibinfo{pages}{249} (\bibinfo{year}{2004}).

\bibitem[{\citenamefont{Balka et~al.}(1988)}]{Balka:1987ty}
\bibinfo{author}{\bibfnamefont{L.}~\bibnamefont{Balka}} \bibnamefont{{\it et~al.}},
  \bibinfo{journal}{Nucl. Instrum. Methods A} \textbf{\bibinfo{volume}{267}},
  \bibinfo{pages}{272} (\bibinfo{year}{1988}).

\bibitem[{\citenamefont{Bertolucci et~al.}(1988)}]{Bertolucci:1987zn}
\bibinfo{author}{\bibfnamefont{S.}~\bibnamefont{Bertolucci}}
  \bibnamefont{{\it et~al.}}, \bibinfo{journal}{Nucl. Instrum. Methods A}
  \textbf{\bibinfo{volume}{267}}, \bibinfo{pages}{301} (\bibinfo{year}{1988}).

\bibitem[{\citenamefont{Albrow et~al.}(2002)}]{Albrow:2001jw}
\bibinfo{author}{\bibfnamefont{M.~G.} \bibnamefont{Albrow}}
  \bibnamefont{{\it et~al.}}, \bibinfo{journal}{Nucl. Instrum. Methods A}
  \textbf{\bibinfo{volume}{480}}, \bibinfo{pages}{524} (\bibinfo{year}{2002}).

\bibitem[{\citenamefont{Breccia et~al.}(2004)}]{Breccia:2004}
\bibinfo{author}{\bibfnamefont{L.} \bibnamefont{Breccia}}
  \bibnamefont{{\it et~al.}}, \bibinfo{journal}{Nucl. Instrum. Methods A}
  \textbf{\bibinfo{volume}{532}}, \bibinfo{pages}{575} (\bibinfo{year}{2004}).

\bibitem[{\citenamefont{Abe et~al.}(1992)}]{Abe:1991ui}
\bibinfo{author}{\bibfnamefont{F.}~\bibnamefont{Abe}} \bibnamefont{{\it et~al.}}
  (\bibinfo{collaboration}{CDF Collaboration}), \bibinfo{journal}{Phys. Rev. D}
  \textbf{\bibinfo{volume}{45}}, \bibinfo{pages}{1448} (\bibinfo{year}{1992}).

\bibitem[{\citenamefont{Bhatti et~al.}(2006)}]{Bhatti:2005ai}
\bibinfo{author}{\bibfnamefont{A.}~\bibnamefont{Bhatti}} \bibnamefont{{\it et~al.}},
  \bibinfo{journal}{Nucl. Instrum. Methods A} \textbf{\bibinfo{volume}{566}},
  \bibinfo{pages}{375} (\bibinfo{year}{2006}).

\bibitem[{\citenamefont{Ascoli et~al.}(1988)}]{Ascoli:1987av}
\bibinfo{author}{\bibfnamefont{G.}~\bibnamefont{Ascoli}} \bibnamefont{{\it et~al.}},
  \bibinfo{journal}{Nucl. Instrum. Methods A} \textbf{\bibinfo{volume}{268}},
  \bibinfo{pages}{33} (\bibinfo{year}{1988}).

\bibitem[{\citenamefont{Dorigo}(2001)}]{Dorigo:2000ip}
\bibinfo{author}{\bibfnamefont{T.}~\bibnamefont{Dorigo}}
  (\bibinfo{collaboration}{CDF Collaboration}), \bibinfo{journal}{Nucl.
  Instrum. Methods A} \textbf{\bibinfo{volume}{461}}, \bibinfo{pages}{560}
  (\bibinfo{year}{2001}).

\bibitem[{\citenamefont{Thomson et~al.}(2002)}]{Thomson:2002xp}
\bibinfo{author}{\bibfnamefont{E.~J.} \bibnamefont{Thomson}}
  \bibnamefont{{\it et~al.}}, \bibinfo{journal}{IEEE Trans. Nucl. Sci.}
  \textbf{\bibinfo{volume}{49}}, \bibinfo{pages}{1063} (\bibinfo{year}{2002}).

\bibitem[{\citenamefont{buzatu}(2012)\citenamefont{Buzatu,
  Warburton, Krumnack, and Yao}}]{BuzatuThesis}
\bibinfo{author}{\bibfnamefont{A.} \bibnamefont{Buzatu}},
  \bibinfo{author}{\bibfnamefont{A.}~\bibnamefont{Warburton}},
  \bibinfo{author}{\bibfnamefont{N.}~\bibnamefont{Krumnack}}, \bibnamefont{and}
  \bibinfo{author}{\bibfnamefont{W.-M.} \bibnamefont{Yao}},
  \bibinfo{journal}{submitted to Nucl. Instrum. Methods A}, \eprint{arXiv:1206.4813}.

\bibitem[{\citenamefont{Acosta et~al.}(2005)}]{Acosta:2004hw}
\bibinfo{author}{\bibfnamefont{D.~E.} \bibnamefont{Acosta}}
  \bibnamefont{{\it et~al.}} (\bibinfo{collaboration}{CDF Collaboration}), \bibinfo{journal}{Phys.
  Rev. D} \textbf{\bibinfo{volume}{71}}, \bibinfo{pages}{052003}
  (\bibinfo{year}{2005}).

\bibitem[{\citenamefont{Abulencia et~al.}(2006)}]{Abulencia:2006kv}
\bibinfo{author}{\bibfnamefont{A.}~\bibnamefont{Abulencia}}
  \bibnamefont{{\it et~al.}} (\bibinfo{collaboration}{CDF Collaboration}), \bibinfo{journal}{Phys.
  Rev. D} \textbf{\bibinfo{volume}{74}}, \bibinfo{pages}{072006}
  (\bibinfo{year}{2006}).

\bibitem[{\citenamefont{Keung}(2010)}]{roma}
\bibinfo{author}{\bibfnamefont{J.}~\bibnamefont{Keung}}
  , \bibinfo{journal}{Ph.D. Thesis, University of Pennsylvania},
  \bibinfo{journal}{FERMILAB-THESIS-2010-32}, (\bibinfo{year}{2010}).

\bibitem[{\citenamefont{Sforza}(2011)}]{svm}
\bibinfo{author}{\bibfnamefont{F.}~\bibnamefont{Sforza}}
 \bibnamefont{{\it et~al.}} \bibinfo{journal}{J. Phys. Conf. Ser.}, 
  \textbf{\bibinfo{volume}{331}}, \bibinfo{pages}{032045} (\bibinfo{year}{2011}).

\bibitem[{\citenamefont{Aaltonen et~al.}(2010)}]{singletop}
\bibinfo{author}{\bibfnamefont{T.}~\bibnamefont{Aaltonen}}
  \bibnamefont{{\it et~al.}} (\bibinfo{collaboration}{CDF Collaboration}), \bibinfo{journal}{Phys.
  Rev. D} \textbf{\bibinfo{volume}{82}}, \bibinfo{pages}{112005}
  (\bibinfo{year}{2010}).


\bibitem[{\citenamefont{Campbell et~al.}(2003)\citenamefont{Campbell, Ellis}}]{Campbell:2002tg}
\bibinfo{author}{\bibfnamefont{J.} \bibnamefont{Campbell}},
  \bibinfo{author}{\bibfnamefont{R.~K.}~\bibnamefont{Ellis}},
  \bibinfo{journal}{Phys. Rev. D} \textbf{\bibinfo{volume}{65}}, \bibinfo{pages}{113007}
  (\bibinfo{year}{2002}).

\bibitem[{\citenamefont{Cacciari et~al.}(2004)\citenamefont{Cacciari, Frixione, 
Mangano, Nason, and Ridolfi}}]{Cacciari:2003fi}
\bibinfo{author}{\bibfnamefont{M.} \bibnamefont{Cacciari}},
  \bibinfo{author}{\bibfnamefont{S.}~\bibnamefont{Frixione}},
  \bibinfo{author}{\bibfnamefont{M.~L.}~\bibnamefont{Mangano}},
  \bibinfo{author}{\bibfnamefont{P.}~\bibnamefont{Nason}}, \bibnamefont{and}
  \bibinfo{author}{\bibfnamefont{G.}~\bibnamefont{Ridolfi}},
  \bibinfo{journal}{J. High Energy Phys.} \bibinfo{volume}{04} (\bibinfo{year}{2004})
\bibinfo{pages}{068}.


\bibitem[{\citenamefont{Harris et~al.}(2002)\citenamefont{Harris, Laenen, 
Phaf, Sullivan, and Weinzierl}}]{Harris:2002md}
\bibinfo{author}{\bibfnamefont{B.~W.} \bibnamefont{Harris}},
  \bibinfo{author}{\bibfnamefont{E.}~\bibnamefont{Laenen}},
  \bibinfo{author}{\bibfnamefont{L.}~\bibnamefont{Phaf}},
  \bibinfo{author}{\bibfnamefont{Z.}~\bibnamefont{Sullivan}}, \bibnamefont{and}
  \bibinfo{author}{\bibfnamefont{S.}~\bibnamefont{Weinzierl}},
  \bibinfo{journal}{Phys. Rev. D} \textbf{\bibinfo{volume}{66}}, \bibinfo{pages}{054024}
  (\bibinfo{year}{2002}).

\bibitem[{\citenamefont{Acosta et~al.}(2005)}]{Acosta:2004uq}
\bibinfo{author}{\bibfnamefont{D.}~\bibnamefont{Acosta}} \bibnamefont{{\it et~al.}}
  (\bibinfo{collaboration}{CDF Collaboration}), \bibinfo{journal}{Phys. Rev. Lett.}
  \textbf{\bibinfo{volume}{94}}, \bibinfo{pages}{091803}
  (\bibinfo{year}{2005}).


\bibitem[{\citenamefont{Mangano et~al.}(2003)\citenamefont{Mangano, Moretti,
  Piccinini, Pittau, and Polosa}}]{Mangano:2002ea}
\bibinfo{author}{\bibfnamefont{M.~L.} \bibnamefont{Mangano}},
  \bibinfo{author}{\bibfnamefont{M.}~\bibnamefont{Moretti}},
  \bibinfo{author}{\bibfnamefont{F.}~\bibnamefont{Piccinini}},
  \bibinfo{author}{\bibfnamefont{R.}~\bibnamefont{Pittau}}, \bibnamefont{and}
  \bibinfo{author}{\bibfnamefont{A.~D.} \bibnamefont{Polosa}},
  \bibinfo{journal}{J. High Energy Phys.} \bibinfo{volume}{07} (\bibinfo{year}{2003}) \bibinfo{pages}{001}.

\bibitem[{\citenamefont{Corcella et~al.}(2001)}]{Corcella:2001wc}
\bibinfo{author}{\bibfnamefont{G.}~\bibnamefont{Corcella}} \bibnamefont{{\it et~al.}},
 \eprint{arXiv:hep-ph/0201201}.

\bibitem[{\citenamefont{Sjostrand et~al.}(2001)}]{Sjostrand:2000wi}
\bibinfo{author}{\bibfnamefont{T.}~\bibnamefont{Sjostrand}}
  \bibnamefont{{\it et~al.}}, \bibinfo{journal}{Comput. Phys. Commun.}
  \textbf{\bibinfo{volume}{135}}, \bibinfo{pages}{238} (\bibinfo{year}{2001}).


\bibitem[{\citenamefont{Abulencia et~al.}(2006)}]{Abulencia:2005aj}
\bibinfo{author}{\bibfnamefont{A.}~\bibnamefont{Abulencia}}
  \bibnamefont{{\it et~al.}} (\bibinfo{collaboration}{CDF Collaboration}),
  \bibinfo{journal}{Phys. Rev. D} \textbf{\bibinfo{volume}{73}},
  \bibinfo{pages}{032003} (\bibinfo{year}{2006}{\natexlab{d}}).

\bibitem[{\citenamefont{Pumplin et~al.}(2002)}]{Pumplin:2002vw}
\bibinfo{author}{\bibfnamefont{J.}~\bibnamefont{Pumplin}} \bibnamefont{{\it et~al.}},
  \bibinfo{journal}{J. High Energy Phys.} \bibinfo{volume}{07} (\bibinfo{year}{2002}) \bibinfo{pages}{012}.

\bibitem[{\citenamefont{Aaltonen et~al.}(2011)\citenamefont{Aaltonen, Buzatu,
  Kilminster, Nagai, and Yao}}]{Timo_NIM}
\bibinfo{author}{\bibfnamefont{T.} \bibnamefont{Aaltonen}},
  \bibinfo{author}{\bibfnamefont{A.}~\bibnamefont{Buzatu}},
  \bibinfo{author}{\bibfnamefont{B.}~\bibnamefont{Kilminster}},
  \bibinfo{author}{\bibfnamefont{Y.}~\bibnamefont{Nagai}}, \bibnamefont{and}
  \bibinfo{author}{\bibfnamefont{W.-M.} \bibnamefont{Yao}},
  \bibinfo{journal}{submitted to Nucl. Instrum. Methods A}, \eprint{arXiv:1107.3026}.

\bibitem[{\citenamefont{Junk}(1999)}]{trj}
\bibinfo{author}{\bibfnamefont{T.} \bibnamefont{Junk}},
\bibinfo{journal}{Nucl. Instrum. Methods A}
  \textbf{\bibinfo{volume}{434}}, \bibinfo{pages}{435}
  (\bibinfo{year}{1999}).

\end{thebibliography}
%\input{PRD_WH7.5fb.bbl}

\end{document}